\documentclass[aps,prl,preprintnumbers,twocolumn,superscriptaddress,nofootinbib,floatfix,10pt]{revtex4-1}
\usepackage{amsfonts,amssymb,stmaryrd,latexsym,amsmath}
\usepackage{graphicx,subfigure}
\usepackage{times}
\usepackage{slashed}
\usepackage{braket}
\usepackage{verbatim}
\usepackage{enumitem}
\usepackage[utf8]{inputenc}
\usepackage[nocomma]{optidef}
\usepackage{eqnarray}

\usepackage{url}
\makeatletter
\g@addto@macro{\UrlBreaks}{\UrlOrds}
\makeatother

\newcounter{Caux}
\newcounter{Cequ}
\newenvironment{Calignat}
{\setcounter{Caux}{\theequation}
\setcounter{equation}{\theCequ}%
\renewcommand\theequation{C.\arabic{equation}}
\alignat}{\endalignat\setcounter{Cequ}{\value{equation}}\setcounter{equation}{\theCaux}}
\usepackage{color}
\usepackage{hyperref}

\usepackage[normalem]{ulem}
\usepackage{comment}

\makeatother
\begin{document}

\title{Qubit-reuse compilation with mid-circuit measurement and reset} 


\author{Matthew DeCross}
\email{matthew.decross@quantinuum.com}
\affiliation{Quantinuum, 303 S Technology Ct, Broomfield, CO 80021, USA}

\author{Eli Chertkov}
\email{eli.chertkov@quantinuum.com}
\affiliation{Quantinuum, 303 S Technology Ct, Broomfield, CO 80021, USA}

\author{Megan Kohagen}
\email{megan.l.kohagen@quantinuum.com}
\affiliation{Quantinuum, 303 S Technology Ct, Broomfield, CO 80021, USA}

\author{Michael Foss-Feig}
\email{michael.feig@quantinuum.com}
\affiliation{Quantinuum, 303 S Technology Ct, Broomfield, CO 80021, USA}

\begin{abstract}

A number of commercially available quantum computers, such as those based on trapped-ion or superconducting qubits, can now perform mid-circuit measurements and resets.
In addition to being crucial for quantum error correction, this capability can help reduce the number of qubits needed to execute many types of quantum algorithms by measuring qubits as early as possible, resetting them, and reusing them elsewhere in the circuit.
In this work, we introduce the idea of qubit-reuse compilation, which takes as input a quantum circuit and produces as output a compiled circuit that requires fewer qubits to execute due to qubit reuse.
We present two algorithms for performing qubit-reuse compilation: an exact constraint programming optimization model and a greedy heuristic.
We introduce the concept of \emph{dual circuits}, obtained by exchanging state preparations with measurements and vice versa and reversing time, and show that optimal qubit-reuse compilation requires the same number of qubits to execute a circuit as its dual.
We illustrate the performance of these algorithms on a variety of relevant near-term quantum circuits, such as one-dimensional and two-dimensional time-evolution circuits, and numerically benchmark their performance on the quantum adiabatic optimization algorithm (QAOA) applied to the MaxCut problem on random three-regular graphs.
To demonstrate the practical benefit of these techniques, we experimentally realize an 80-qubit QAOA MaxCut circuit on the 20-qubit Quantinuum H1-1 trapped ion quantum processor using qubit-reuse compilation algorithms.
\end{abstract}

\maketitle

\section{Introduction}

Current quantum computers are limited by the number of qubits available for computation. To conclusively demonstrate the computational advantage of these devices compared to their classical counterparts on a variety of practical applications, researchers will need to make efficient use of qubits.

Generally, the process of submitting a quantum circuit to solve a problem of interest consists of devising a quantum algorithm, generating the associated quantum circuit, performing circuit optimization for a quantum device of interest, submitting to the quantum device, and retrieving the results. The step of circuit optimization is key on noisy intermediate-scale quantum (NISQ) devices. Circuit optimization takes many different forms but typically consists of modifying the gate structure of a quantum circuit to increase the fidelity of results. Some examples of circuit optimizations include reducing gate count by simplifying Clifford subcircuits \cite{Bravyi21, PhysRevA.70.052328, Qiskit, Sivarajah_2020}, removing redundant gates \cite{Sivarajah_2020, Jang21, Qiskit}, and using the KAK decomposition of two-qubit gates \cite{Khaneja2000}, or mapping circuits to a specific quantum device architecture to decrease potential transport or swap gate costs \cite{Sivarajah_2020, Qiskit, Lin22, Zulehner17}.

An underexplored area of quantum circuit design is that of qubit reuse. The ability of a quantum computer to perform mid-circuit measurements and resets enables the reuse of qubits after resets. Qubit reuse is an essential ingredient of scalable quantum error correction protocols \cite{RyanAnderson2022}, which require repeated mid-circuit measurements and resets of ancilla qubits to measure error syndromes, and is already available in trapped-ion \cite{Pino:2021} and superconducting \cite{IBMBlogPost} qubit architectures. Recently, qubit-reuse techniques have been used to experimentally prepare and time-evolve large tensor network states on trapped ion quantum computers \cite{Chertkov2022}, to study a non-equilibrium phase transition \cite{chertkov2022b}, and to perform entanglement spectroscopy of quantum systems \cite{Yirka2021qubitefficient}.

In this paper, we investigate the qubit-reuse compilation problem: the task of converting a quantum circuit into an equivalent circuit that uses fewer qubits via qubit reuse. We demonstrate two algorithms for performing qubit-reuse compilation. The first is an exact algorithm that uses a constraint programming and satisfiability (CP-SAT) model. Since this algorithm uses a general-purpose type of model for solving combinatorial optimization problems, it does not scale well with qubit number, but provides a useful benchmark for other methods at low qubit number. The second algorithm is a greedy heuristic that runs quickly up to large numbers of qubits and scales polynomially with qubit number.

We benchmark these algorithms numerically on the quantum approximate optimization algorithm (QAOA) MaxCut circuits on random three-regular graphs. We also investigate several examples of highly structured circuits, including 1D and 2D time evolution circuits and certain quantum tensor networks, and solve the qubit-reuse compilation problem analytically for these cases. Using our algorithms, we are able to compress an eighty-qubit QAOA circuit into a twenty-qubit circuit, which we then experimentally execute on the Quantinuum H1-1 trapped ion quantum computer \cite{H1-1}, which has high-fidelity and low-crosstalk mid-circuit measurement and reset capabilities \cite{Pino:2021}, as well as the recently announced \cite{ArbitraryAngleBlogPost} support for arbitrary angle two-qubit rotation gates.

Techniques for executing large quantum circuits on a smaller number of qubits have been developed previously in, for example, \cite{Lowe22, Li21, Saleem21, Tang22, Peng19, Piveteau22}. These methods are typically based on cutting the original circuits into smaller circuits. Depending on the way in which this is accomplished there can be a number of drawbacks to this approach. For instance, circuits corresponding to graph problems can be cut by approximately factorizing the graph by removing special edges. However, the resulting measurements cannot sample from fully entangled states and only correspond to approximate solutions even in the limit of very deep circuits. Methods based on inserting randomized measurements or Pauli operators circumvent this problem but generally proliferate the number of circuits exponentially in the number of cut wires, or require distributed networks of quantum computers to run the resulting cut circuits.

Our work is novel in that it provides a one-to-one mapping of a quantum circuit into a single equivalent circuit that uses fewer qubits and samples from the identical measurement distribution as the original circuit. The key underlying technique is not based on circuit cutting but rather takes advantage of the causal structure of the circuit, as elucidated below. We also use this technique to execute the full QAOA optimization protocol on hardware, demonstrating the near-term practical application of qubit-reuse compilation.

\section{Qubit Reuse Via Mid-Circuit Measurement}

\begin{figure*}[!t]
    \centering
    \includegraphics[width = .8\textwidth]{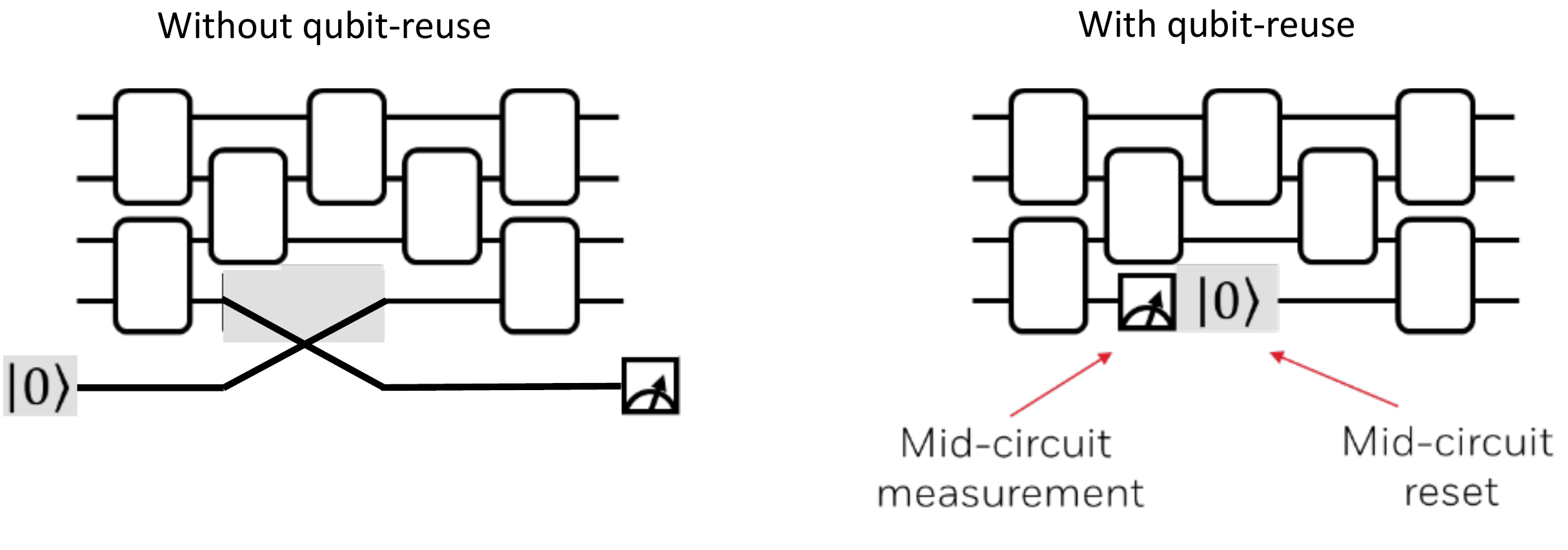}
    \caption{On the left, the lowermost input qubit is not used until after crossing the shaded grey region, while the adjacent qubit is no longer used after that region. The overall circuit can be executed using one fewer qubit using mid-circuit measurement and reset, as depicted on the right.}
    \label{fig:reuse}
\end{figure*}

The key principle underlying the ability to measure and reuse qubits to reduce qubit number overhead is the fact that in many cases only partial execution of a circuit encoding a quantum program is required to measure a given output qubit, at which point that qubit can potentially be reset to play the role of an input qubit elsewhere in the circuit. Fig.~\ref{fig:reuse} demonstrates how an output qubit that is no longer required to implement the remainder of a circuit can be measured, reset, and reused as a different input.

Specifically, a given output qubit's quantum state depends only on gates and qubits in its past \emph{causal cone}. Exploiting the causal structure of quantum circuits in quantum simulation is a technique frequently utilized in classical tensor network methods, such as those based on the multiscale entanglement renormalization ansatz (MERA) \cite{PhysRevLett.101.110501}, and has been recently applied to numerous quantum extensions of tensor-network methods for both many-body physics and machine learning \cite{kim2017,Kim_2017b,Huggins_2019,PhysRevResearch.1.023025,FossFeig2021,PhysRevLett.127.040501,Chertkov_2021,Chertkov_2022,Wall_2022} and to several examples of variational quantum algorithms in \cite{PhysRevResearch.3.033083, Amaro_2022, Linke_2019}.

\begin{figure*}[!t]
    \centering
    \includegraphics[width = .9\textwidth]{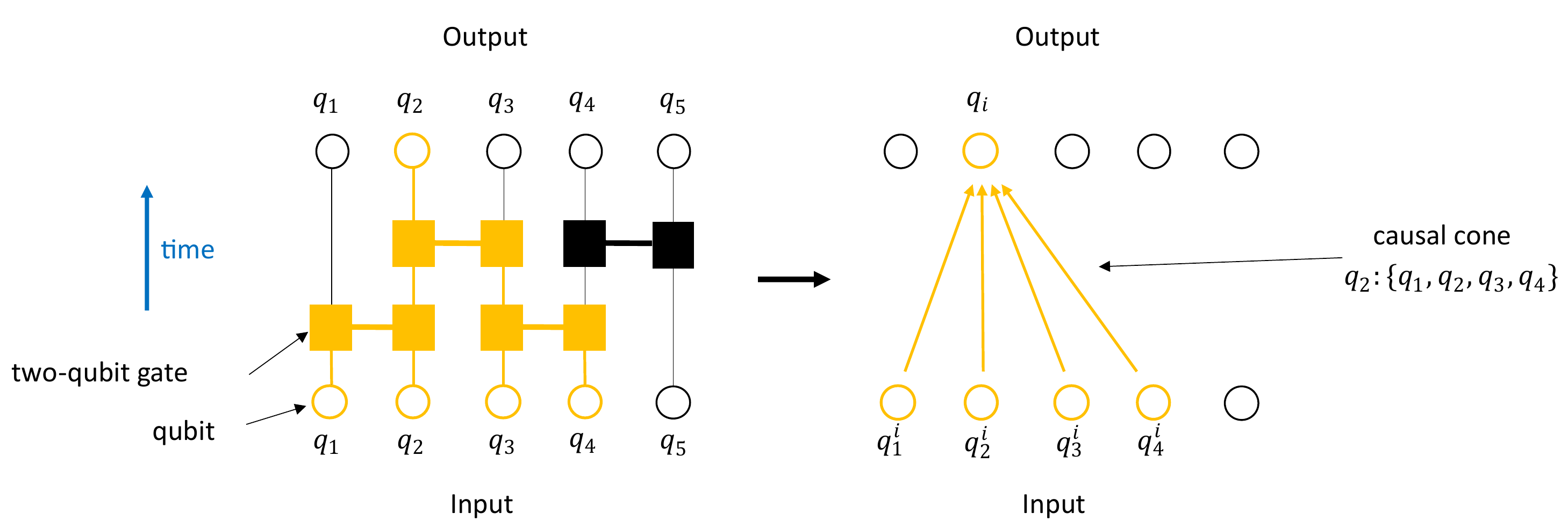}
    \caption{Identification of the causal cone of the output qubit $q_2$ in the displayed quantum circuit. Measuring and resetting $q_2$ only requires gates executed between the four input qubits $\{q_1, q_2, q_3, q_4\}$, after which $q_2$ can be reset and re-used as $q_5$.}
    \label{fig:causal_cones}
\end{figure*}

However, most previous examples of compression based on qubit reuse have proceeded by inspection of circuits with particularly simple structure, without a general framework for automating the compilation of a circuit to run on fewer qubits. In this work, we have developed algorithms that determine an order for measuring and reusing qubits based on their causal cones within a quantum circuit. Fig.~\ref{fig:causal_cones} illustrates how a given output qubit depends causally on a restricted subset of input qubits.

The process of compiling a given quantum circuit into a circuit with fewer qubits is performed in two steps. (1) We determine an order in which to measure and reset the qubits in the original circuit. This order can be determined numerically by the algorithms we describe below, which aim to globally minimize the number of qubits required to run the final circuit, or can be determined analytically. (2) We use the determined measurement order to rewrite the original circuit into a smaller circuit by measuring and resetting the qubits in the desired order. We emphasize that \emph{any} measurement order can be used in (2), so one does not need to restrict oneself to minimizing the qubit number but can instead determine measurement orders that optimize more complicated objectives, e.g., objectives that strike a balance between minimizing hardware errors and minimizing qubits on a noisy qubit-limited device. Both (1) and (2) have been fully automated and integrated into a software package used to produce the numerical results discussed in this work.

We note that a circuit $\mathcal{C}$ and the version of that circuit compressed via qubit reuse, $\mathcal{R}(\mathcal{C})$, do not generate equivalent operators on a set of qubits (they do not even act on qubit sets of the same size).  Rather their equivalence is in the restricted sense that they produce measurement data drawn from identical distributions, which we denote as $\mathcal{R}(\mathcal{C})\cong \mathcal{C}$.  This equivalence is fairly strong, in that no classical post-processing of the circuit output can distinguish $\mathcal{R}(\mathcal{C})$ from $\mathcal{C}$.  However, quantum post-processing of the state generated prior to measurement in $\mathcal{C}$ (i.e. extending the circuit $\mathcal{C}$ to include more gates) cannot generally be carried out in an equivalent fashion on $\mathcal{R}(\mathcal{C})$, since $\mathcal{R}(\mathcal{C})$ never actually produces the quantum state generated by $\mathcal{C}$ immediately prior to measurement.

We also point out that a given quantum circuit $\mathcal{C}$ and its compressed version  $\mathcal{R}(\mathcal{C})$ only contain an identical set of gates in the case that the target device architecture supports interactions between arbitrary qubits, i.e. is fully connected. In architectures that support only nearest-neighbor or similar interactions, both the circuit and its compressed version will generally require the insertion of SWAP gates to execute, and the location and number of these SWAP gates might not be the same in both the circuit and its compressed version.

\section{Algorithms for Optimizing Qubit Reuse}

\subsection{Exact Solution by Constraint Programming}

The problem of minimizing the required number of qubits to execute a given quantum program can be formulated in the language of a constraint programming and satisfiability (CP-SAT) problem. The optimization version of CP-SAT is formulated as the minimization of an objective function subject to a set of equality and inequality constraints as well as (potentially) a set of conditional constraints. By formulating the qubit-reuse compilation problem as a CP-SAT problem, we can use existing solvers to find globally optimal solutions. While in the worst case these solvers run in time superpolynomial in the problem size, they can be used to obtain exact solutions for small quantum circuits, or to find approximate solutions for larger circuits by constraining the run time.

Suppose that we are given a quantum circuit defined on $N$ qubits and wish to determine a measurement order of the $N$ qubits that will allow us to execute the program on (possibly) fewer than $N$ qubits by measuring, resetting, and reusing certain qubits. We define the following problem input data:
\begin{align*}
    q \in Q: \quad &\text{qubits in the circuit}   \\ 
    t \in T: \quad &\text{qubit-reuse time steps in the circuit} \\
    C_{q}: \quad &\text{set of qubits in the causal cone of qubit $q$}
\end{align*}

Here $q \in Q =\{1,\ldots, N\}$ indexes the original set of $N$ qubits $Q$ and $t \in T = \{1, \ldots, N\}$ indexes the $N$ time steps at which each of the original $N$ qubits is measured and reset. At each time step, one qubit is measured and reset. Note that $t$ is not necessarily related to the usual notion of time in quantum circuits, i.e., the number of layers of two-qubit gates applied to the input state. To ``run the circuit up to time $t$" means to apply only the gates in the causal cones of all qubits measured at times $1,2,\ldots,t$.

The optimization model for the qubit-reuse problem is defined in terms of two sets of $N^2$ binary variables and an integer-valued cost function $C$ to be minimized:
\begin{align*}
    m_{qt} \in \{0,1\}: \quad &\text{qubit $q$ is measured at time $t$ or not}   \\ 
    c_{qt} \in \{0,1\}: \quad &\text{qubit $q$ is in a causal cone of any qubit} \\
    & \text{measured at $t^{\prime} \leq t$ and  qubit $q$ has not been}         \\
    & \text{measured yet before time $t$, otherwise not}                         \\
    C \in \mathbb{Z}^+: \quad &\text{maximum number of qubits used in circuit}
\end{align*}

The variable $m_{qt}$ tracks when qubits are measured. The variable $c_{qt}$ keeps track of which qubits are required to execute the circuit up to time $t$. The above descriptions of the variables $m_{qt}, c_{qt}$ and the cost function $C$ are enforced by the constraints in the model. 
 
In words, the constraints are:

 \begin{enumerate}[label = {(C.\arabic*)}, start = 1]

\item The number of qubits required to run the circuit is the maximum number of qubits required at any given time.

\item If qubit $q$ is measured at time $t$, then it is in at least one causal cone of a qubit that has been measured at some $t' \leq t$.

\item If a qubit is required to run the circuit up to time $t-1$, it is either measured at time $t-1$ or was in the causal cone of at least one qubit measured up to time $t$.

\item If qubit $q$ is measured at time $t$, it is in at least one causal cone of a qubit that has been measured up to time $t$.

\item If qubit $q$ is measured at time $t$, then it is no longer needed to run the circuit at subsequent times.\footnote{This does not mean the circuit is independent of $q$ at later times; rather, it means that the operations involving $q$ have already been accounted for.}

\item Each qubit is measured once.

\item Only one qubit is measured at each time.
    
\end{enumerate}

\noindent The technical statement of each of these constraints can be found in the Supplemental Materials. We emphasize that without the constraints, the binary variables $m_{qt}$ and $c_{qt}$ have no meaning; the purpose of the constraints is to provide a set of equality, inequality, and conditional constraints on these variables that correspond to the physical problem of interest. For example, constraint (C.1) defines the cost function $C$ in terms of a certain sum of $c_{qt}$ variables that corresponds to the number of qubits you need to implement the circuit up to a certain time. The cost function of interest is the number of qubits you need to run the circuit, which in turn is the size of the worst-case (maximum) number of qubits needed to run the circuit at any time.

We solve the model specified by minimization of the objective function subject to the constraints \eqref{eq:first_constraint} - \eqref{eq:last_constraint} using the CP-SAT solver provided by Google's open-source OR-Tools package \cite{ortools}. This solver employs constraint programming techniques combined with SAT solving techniques to solve the problem exactly and is well-suited for problems with binary variables \cite{Stuckey2013}. It is also possible for the solver to return the best value of the objective found respecting the constraints within a given time limit, and to seed the solver with a ``hint" solving the constraints at a non-optimal value of the objective. However, our numerical experiments indicate that for this particular problem the search space is sufficiently large that typically either a) the solver produces an exact solution or b) the solver cannot find a feasible solution or improve upon a hint in a reasonable amount of time, depending on the size of the problem instance. Using the output of the CP-SAT model, it is straightforward to extract the prescribed measurement order of the output qubits by examination of the variables $m_{qt}$: at time step $t$ (index $t$ in the measurement order), measure the unique qubit $q$ for which $m_{qt} = 1$.

\subsection{A Local Greedy Heuristic Algorithm}

Although the CP-SAT model furnishes an exact solution to the qubit-reuse compilation problem, it scales poorly with qubit number as it involves $\mathcal{O}(N^2)$ variables and $\mathcal{O}(N^4)$ constraints with nontrivial structure, which becomes prohibitive in the $N \approx 100 - 1000$ qubit regime that will be a near-term proving ground for quantum computers. For practical purposes, it is therefore important to develop heuristic algorithms that can produce useful approximate solutions in a reasonable amount of time. In this section, we describe a local greedy algorithm that accomplishes this. Later, we will show that this greedy heuristic is extremely effective (and often exact) for many circuit structures of near-term relevance.

Using the variables defined in the previous section, the local greedy algorithm proceeds as follows:
\begin{enumerate}

\item First, measure the qubit $q$ with the fewest number of input qubits in its causal cone $C_q$.

\item Next, measure the qubit $q'$ whose causal cone $C_{q'}$ adds the fewest \emph{new} input qubits to $C_q$.

\item Repeat step (2), successively choosing the next qubit $q''$ to measure at time $t$ as the qubit whose causal cone $C_{q''}$ adds the fewest new input qubits to the union of all causal cones of qubits measured so far.
    
\end{enumerate}

A simple improvement of this heuristic algorithm that we also study employs a brute force search over the possible choices of first qubit, at the expense of a multiplicative $\mathcal{O}(N)$ time complexity. That is, instead of first choosing the qubit with the fewest number of input qubits in its causal cone, consider all possible choices of initial qubit. For each possible initial qubit, use steps (2) and (3) above to iteratively construct the full measurement order. Then, select the ordering that required the fewest number of total qubits to execute the program. As numerical results substantiate, this small modification significantly improves the amount of compression achieved by qubit reuse with only a modest performance penalty. Though we do not explore it in this work, one could also perform a brute-force search over the first $k$ qubits in the measurement order, which would multiply the run-time overhead by $\mathcal{O}(N^k)$. 

\subsection{Dual Circuits}

Consider any circuit $\mathcal{C}$ that involves quantum gates, state preparations, and measurements, where the state preparations and measurements can be interspersed arbitrarily within the circuit (qubits can also be traced out at the end of the circuit, which we will regard as measurement with the result discarded).   A circuit $\mathcal{C}^{\star}$ that can be written by replacing all state preparations in $\mathcal{C}$ with measurements and all measurements in $\mathcal{C}$ by state preparations, and then reading the circuit from right to left, will be referred to as \emph{dual} to $\mathcal{C}$.

\begin{figure}[!ht]
    \centering
    \includegraphics[width=\columnwidth]{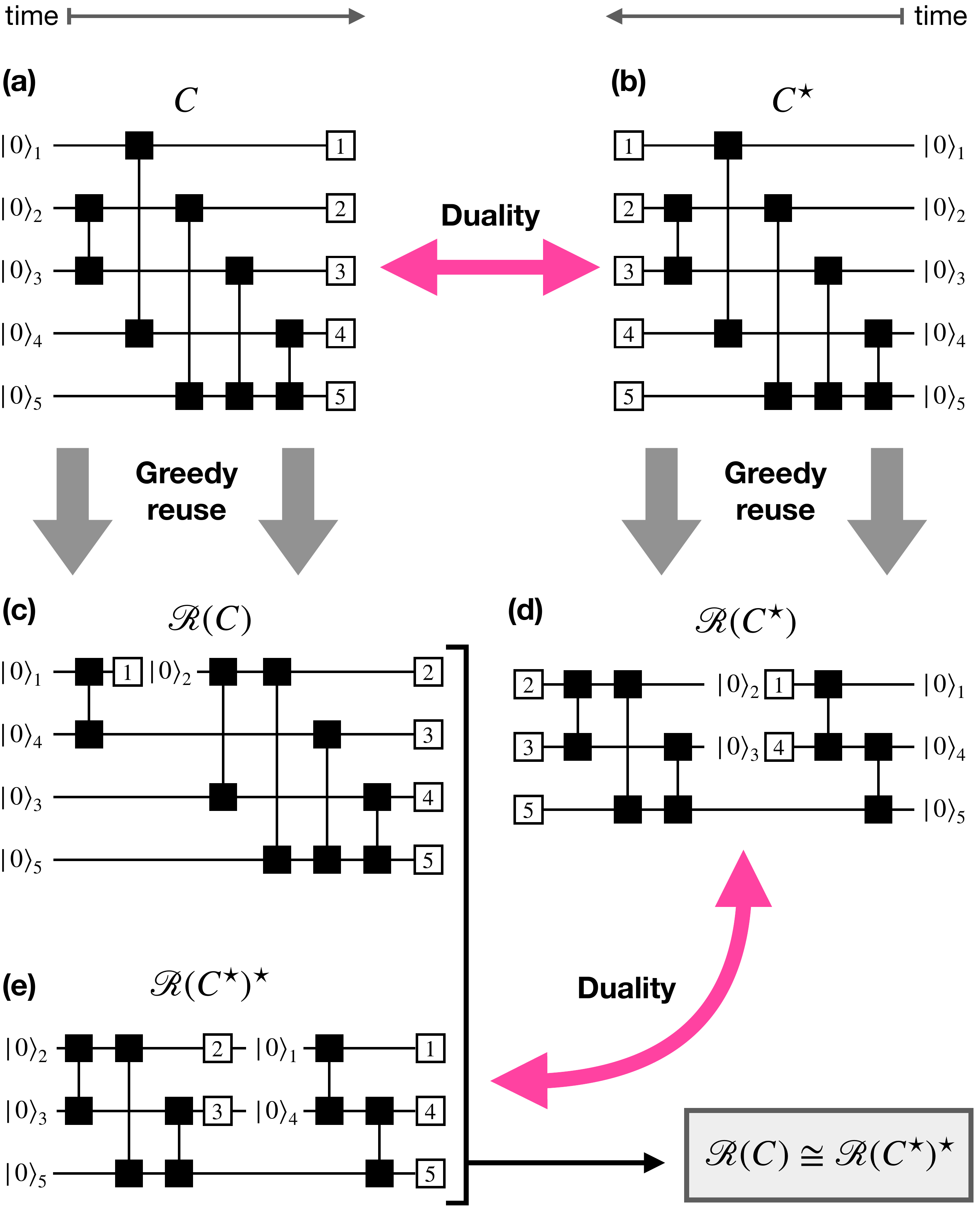}
    \caption{(a) A five-qubit quantum circuit $C$. (b) The dual circuit $C^\star$ of circuit $C$, with time flowing in the opposite direction and with exchanged measurements and resets. (c) The compressed version $\mathcal{R}(C)$ of circuit $C$, after applying the greedy qubit-reuse heuristic, which involves four qubits. (d) The (greedily) compressed version $\mathcal{R}(C^\star)$ of circuit $C^\star$. (e) The dual circuit of $\mathcal{R}(C^\star)$, which involves only three qubits as opposed to the four in $\mathcal{R}(C)$ (see (c)).}
    \label{fig:dual_circuits}
\end{figure}

The compression of $\mathcal{C}$ via qubit reuse into $\mathcal{R}(\mathcal{C})$ can be understood as the following procedure: (1) move all state preparation operations as late in time as possible, (2) move all measurement operations as early in time as possible, (3) for each measurement, optionally connect the associated qubit wire to a state preparation occurring later in time, thus reducing the number of required qubits. Note that the only rule one must obey in this procedure is to never move two operations past each other in time if they have shared support. Performing this compression from $\mathcal{C}$ to $\mathcal{R}(\mathcal{C})$ using the greedy heuristic is shown as Fig.\,\ref{fig:dual_circuits}(a$\rightarrow$ c).  We can also compress the dual circuit $\mathcal{C}^{\star}$ into $\mathcal{R}(\mathcal{C}^{\star})$ [Fig.\,\ref{fig:dual_circuits}(b$\rightarrow$d)], at which point we are free to replace all measurements and state preparations in $\mathcal{R}(\mathcal{C}^{\star})$ according to the replacements mapping $\mathcal{C}^{\star}\rightarrow\mathcal{C}$ in order to generate a new circuit $\mathcal{R}(\mathcal{C}^{\star})^{\star}$ [Fig.\,\ref{fig:dual_circuits}(e)].  Note that $\mathcal{R}(\mathcal{C}^{\star})^{\star}$ \emph{could have been} obtained directly from $\mathcal{C}$ by obeying the compression rules, since respecting time-ordering allows all the same rearrangements and rewiring for either a circuit or its dual.  As a result, we have the general relation

\begin{align*}
\mathcal{R}(\mathcal{C})\cong\mathcal{R}(\mathcal{C}^{\star})^{\star}.
\end{align*}

\begin{figure*}[!t]
    \centering
    \includegraphics[width=0.9\textwidth]{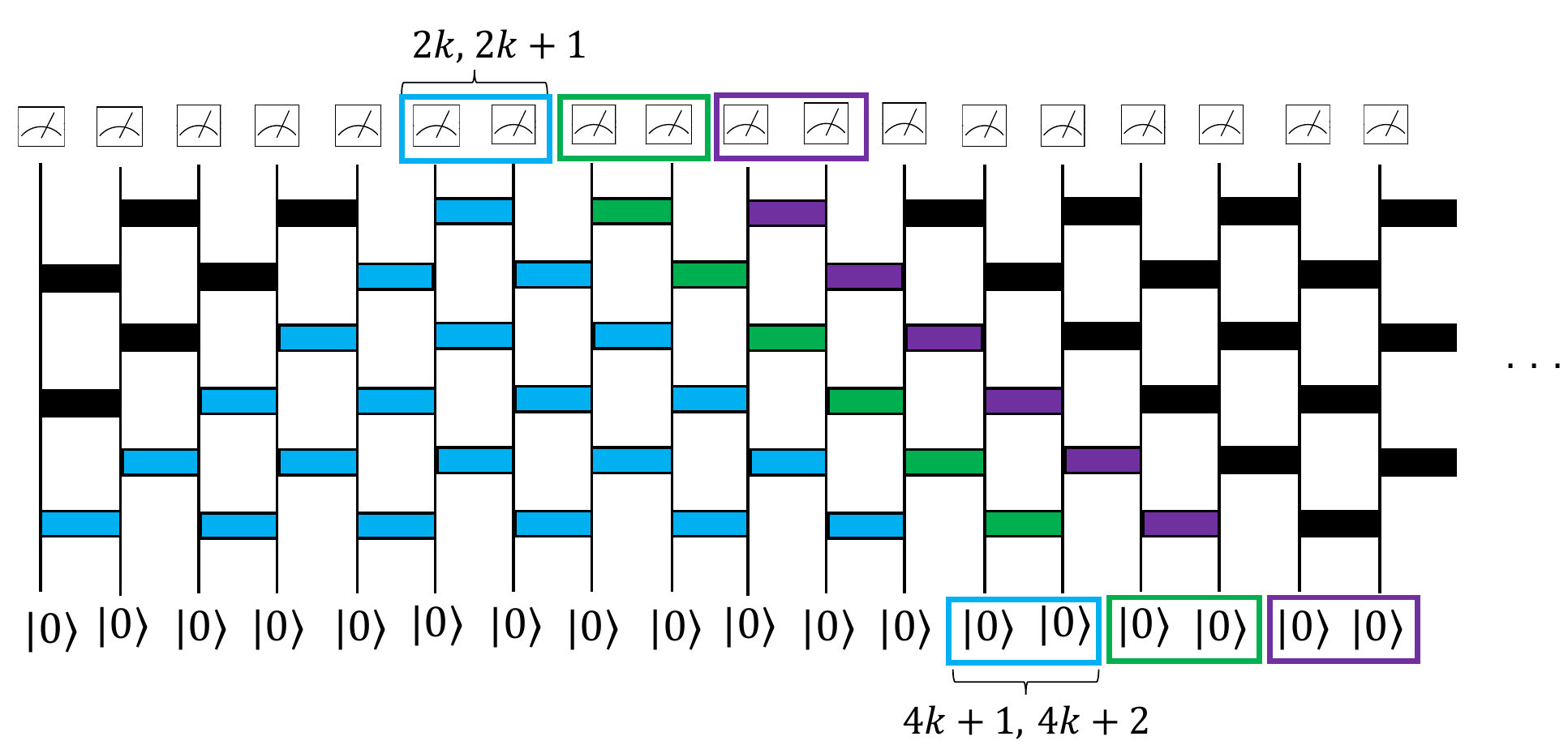}
    \caption{Qubit reuse in one-dimensional brickwork circuit with $k=3$ layers, where time flows upwards by convention. The output qubits labeled $2k, 2k+1$ (blue) are measured first, and reused as input qubits $4k+1, 4k+2$. Those input qubits are subsequently used to implement the green causal cone and measure the green output qubits $2k+2, 2k+3$, which are reused as the purple input qubits, and so on.
    }
    \label{fig:1d_brickwork}
\end{figure*}

Note that this equivalence implies that the optimally compressed version of $\mathcal{C}$ must always contain the same number of qubits as the optimally compressed version of $\mathcal{C}^{\star}$.  However, non-optimal techniques, such as the greedy heuristic described above, need not necessarily return the same level of compression on a circuit and its dual, as evidenced by comparing Figs.\,\ref{fig:dual_circuits}(c,d).  This implies an immediate optimization for any heuristic: One should always apply it to both the circuit in question and its dual, and if its dual is more compressible, one should accept $\mathcal{R}(\mathcal{C}^{\star})^{\star}$ [Fig.\,\ref{fig:dual_circuits}(e)] over the less efficient direct compression $\mathcal{R}(\mathcal{C})$ [Fig.\,\ref{fig:dual_circuits}(c)].

\section{Analytic Results for Qubit-Reuse Compilation}

For certain classes of circuits, it is possible to analytically write down measurement orders that achieve significant compression of the circuit by exploiting its underlying structure and symmetries. We emphasize that these measurement orders can in principle achieve better compression than the approximate numerical algorithms discussed above in certain cases. For several of the circuits discussed below, the analytical measurement order is identical to that produced by the local greedy heuristic and can be shown to be optimal. This demonstrates that in many cases of practical interest the greedy heuristic is exact, and quantifies the scaling (with qubit number) of the compression it achieves.

\subsection{Local Brickwork Circuits in 1D}

The first class of circuits we will consider are one-dimensional brickwork circuits, composed of $k$ alternating even-odd layers of two-qubit gates acting with periodic boundary conditions on $N$ qubits, shown in Fig.~\ref{fig:1d_brickwork}. These circuits appear naturally when simulating local 1D Hamiltonians, and qubit reuse in such linear circuits has been explored previously in the context of quantum matrix product states \cite{kim2017,FossFeig2021,Barratt2021,Niu2021,Zhang2022} and their time evolution \cite{Chertkov2022}.

For this particular circuit, the optimal measurement order is the simple linear order $m_{qt}= \delta_{q,(q_0 + t)\textrm{ mod }N}$ with an arbitrary initial qubit $q_0$. After $k$ layers (where here each ``layer" includes successive application of both a row of ``even" gates and a row of ``odd" gates), each qubit's causal cone is of size $4k$, as illustrated in Fig.~\ref{fig:1d_brickwork} for $k=3$. For example, the causal cone of qubit $q=2k$ includes input qubits $1,\ldots, 4k$. Consequently, this measurement order requires $4k$ qubits to execute, which does not scale with $N$. For the linear measurement order with $q_0=2k$, we first measure qubits $q=2k$ and $q=2k+1$, which participate in the same final gate, then reset and re-use them as input qubits $4k+1,4k+2$. Then, we measure qubits $q=2k+2$ and $q=2k+3$, reset and re-use them as input qubits $4k+3,4k+4$, and proceed this way until we measure qubits $q=2k-2,2k-1$. Note that we assume that $4k < N$, since otherwise each qubit's causal cone spans the entire chain and it is not possible to re-use qubits. We assume periodicity in this section to enforce translation symmetry, which allows the initial qubit $q_0$ to be chosen arbitrarily. However, there is no substantial change in the result by relaxing this symmetry; the initial qubit in this case must be chosen at one edge of the linear chain.

In general, for the one-dimensional brickwork circuits, there are many degenerate measurement orders that produce the same result of $4k$ qubits required to execute the compressed circuit. This is because for any pair of qubits $i, i+1$ that are measured, the next causal cone can be chosen to measure $i+2, i+3$ or $i-1, i-2$, proceeding to the right or left down the linear chain respectively. We expect that this choice can have an impact on the memory errors incurred on qubits which have not yet been measured. For example, proceeding linearly down the chain results in qubits at the beginning remaining live for the entire duration of the circuit, whereas alternating back and forth measuring qubit pairs on either side of the initial pair avoids this but slightly increases the average active time of each qubit. Examining the consequences of this trade-off is a subject of future study.

\begin{figure}[!t]
    \centering
    \includegraphics[width=.7\columnwidth]{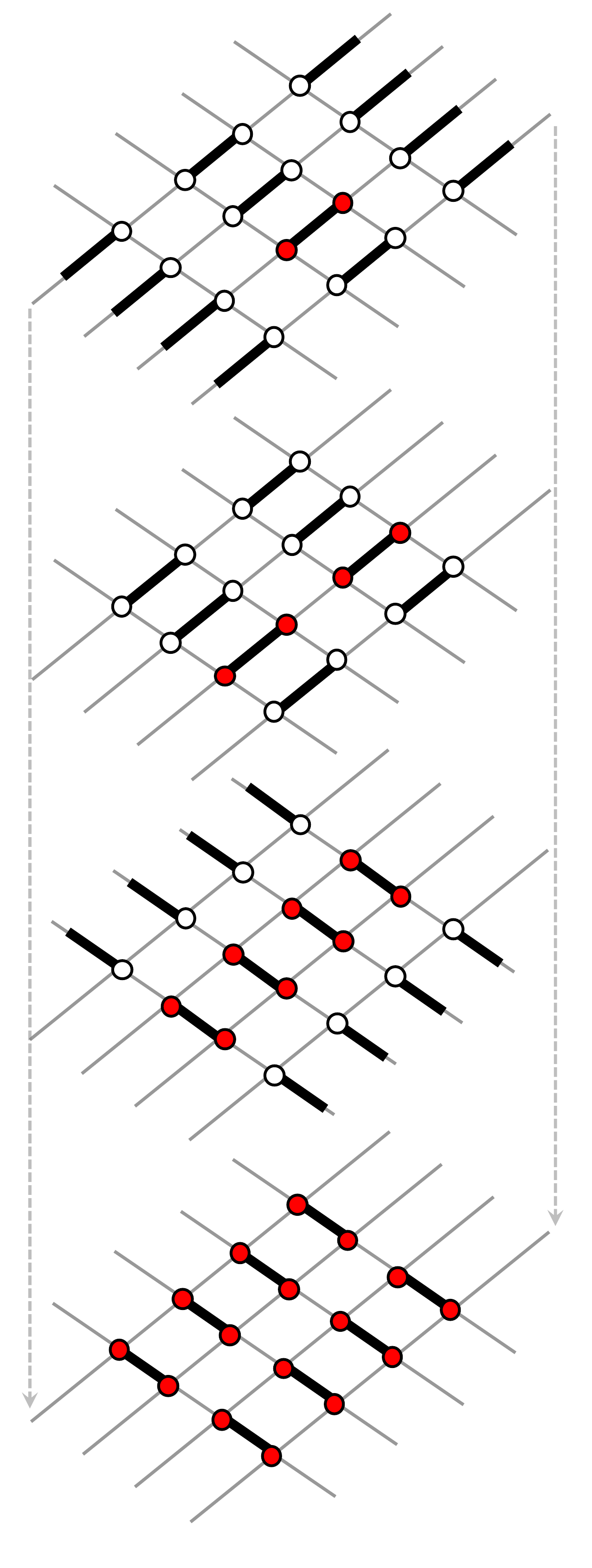}
    \caption{The causal cone of a single qubit after $k=1$ layer of local gates is applied in two dimensions. Read with time flowing from top to bottom, the red coloring of the qubits indicates how causal influence propagates between qubits after each row/column of two-qubit gates is applied. After $k$ layers, the causal cone of the initial qubit has expanded to a size of $16k^2$ at the bottom.}
    \label{fig:2d_layers}
\end{figure}

\subsection{Local Brickwork Circuits in 2D}

It is instructive to understand how the compression of local brickwork circuits generalizes to interactions in higher dimensions. We now examine two-dimensional local brickwork circuits, that is, brickwork circuits where gates are applied between adjacent qubits laid out on a square grid. The analysis in two dimensions demonstrates a general result that $\mathcal{O}(N^{d-1})$ qubits are required to execute a circuit that is originally defined on $\mathcal{O}(N^d)$ qubits in $d$ dimensions, which becomes substantially harder to visualize and analyze in $d > 2$.

Consider a two-dimensional brickwork circuit composed of $k$ layers of gates staggered horizontally and vertically with periodic boundary conditions on an $N \times N$ grid of qubits as shown in Fig.~\ref{fig:2d_layers}. Here, as in the one-dimensional case, a ``layer" refers to the $2d = 4$ total rows of odd+even gates in each dimension. For this circuit, the causal cone of each output qubit can be interpreted as a pyramid containing $4k \times 4k=16k^2$ input qubits (Fig.~\ref{fig:2d_layers}). Here we describe a particular measurement order that performs the optimal qubit reuse for the 2D brickwork circuit. For simplicity, we assume that $N$ is even, and that $4k < N$; we will explain how to generalize the result to non-square lattices without this bound on $k$ at the end.

\begin{figure*}[!t]
    \centering
    \includegraphics[width=\textwidth]{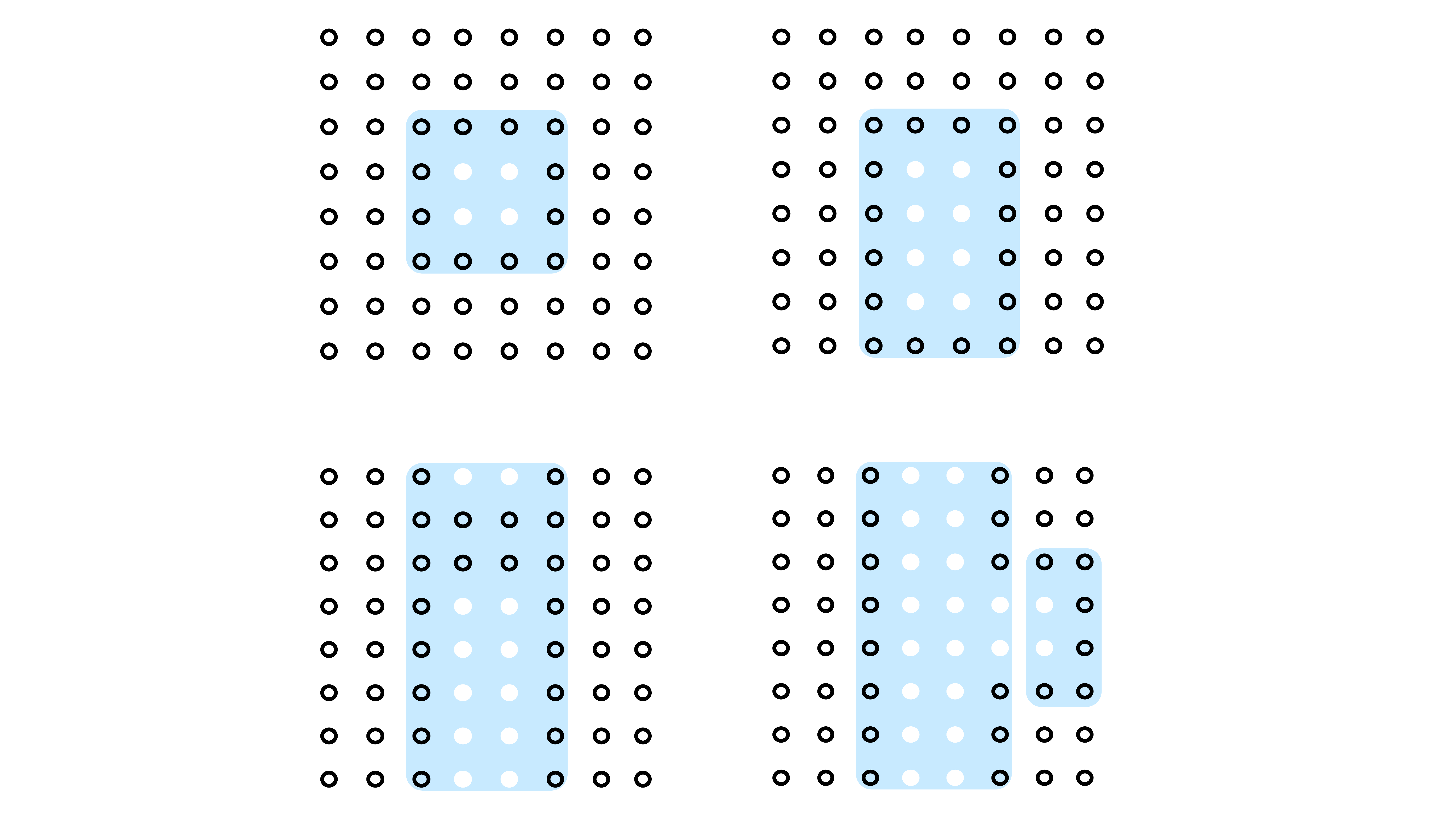}
    \caption{(Top left) The 16 input qubits in the shared causal cone (light blue) of the four white qubits, for $k=1$ layer of gates. (Top right) The greedy algorithm dictates that the next four output qubits to be measured are the four white qubits directly below the first four. Their causal cone extends the required set of input qubits downwards by two (light blue). (Bottom left) The four qubits in the middle of the last causal cone in a complete column can be measured for free without any additional qubits. (Bottom right) Extending to the next column over requires eight new qubits (light blue, right-hand side), but eight qubits are available for reuse from the previous column.}
    \label{fig:2d_reuse}
\end{figure*}

Again, due to translational symmetry, the first qubit $q_0$ to be measured is arbitrary. Furthermore, it is straightforward to check that the qubits on the lattice can be partitioned into groups of four that all share the same causal cone (Fig.~\ref{fig:2d_reuse}). Consequently, a greedy approach will again be optimal, since adjacent qubits either share identical causal cones or their causal cones are simply geometrically shifted along the lattice and heavily overlapping. The most straightforward strategy to count the number of qubits required to implement the greedy algorithm in this case is to tile the lattice column-by-column, by measuring blocks of four adjacent input qubits. After completing a column, one measures a block of four output qubits in the adjacent column, and proceeds to tile that column. This process repeats until the entire lattice has been measured; see Fig.~\ref{fig:2d_reuse} for a clarifying depiction.

We now proceed to count the number of qubits required to implement the brickwork circuit on the 2D lattice according to the process detailed above. As discussed, the first causal cone requires $16k^2$ input qubits to measure the first four output qubits (Fig.~\ref{fig:2d_reuse}, top left). The next causal cone reclaims another four output qubits and requires an additional $8k$ inputs, of which four are reused from the previous step (Fig.~\ref{fig:2d_reuse}, top right). This process repeats until an entire column has been tiled by causal cones, and the last sets of four output qubits in the column are obtained for free since their causal cones fully overlap with causal cones of previously measured qubits (Fig.~\ref{fig:2d_reuse}, bottom left). After the very first causal cone, one requires $N/2 - 2k$ additional causal cones to tile a single column before the subsequent causal cones are completely overlapping. Consequently, the last $2k - 1$ causal cones each reclaim four qubits for free, which yields $4(2k-1) + 4 = 8k$ qubits available for reuse in the next column after completing a single column, where the extra $4$ comes from the last causal cone before overlaps occur.

In total, therefore, measuring the $2N$ output qubits corresponding to a single column of causal cones requires the following number of simultaneously active input qubits:
\begin{align}
    16k^2 + (8k - 4) (N/2 - 2k) = (4k-2)N + 8k \label{eq:2d_reqd}
\end{align}

When we proceed to the next column and repeat this process, the next causal cone requires $8k$ input qubits (Fig.~\ref{fig:2d_reuse}, bottom right). This conveniently exactly matches the number of qubits reclaimed for reuse at the end of the previous column, so no new qubits are required. Furthermore, this causal cone reclaims four output qubits, and each subsequent causal cone in the new column requires only four additional input qubits, not $8k$ (see Fig.~\ref{fig:2d_reuse}, bottom right - the reason is that the causal cones now overlap on two sides). Therefore, the entire column can be measured without use of any additional input qubits, as can all subsequent columns. The last few columns will overlap, but this can only reduce the required overhead.

In summary, therefore, the number of qubits required to execute the circuit containing only the causal cones that tile the first column is the same as the number of qubits required to execute the entire circuit. This number is, from \eqref{eq:2d_reqd}, $(4k-2) N + 8k$.

In the slightly more general case that the lattice is not square and is instead $N_x \times N_y$, the same argument applies where $N$ in \eqref{eq:2d_reqd} should be replaced with the dimension that you first choose to tile with causal cones. To minimize the overhead, therefore, you should choose the shorter dimension first, replacing $N$ with $\min (N_x, N_y)$ above.

In the case that $4k > \min(N_x, N_y)$, the counting arguments above are slightly modified since the size of each causal cone already spans one or more dimensions. Performing the same counting procedure taking this into account yields the following general formula for the compiled qubit number:
\begin{align*}
    \begin{cases}
    (4k-2) \min(N_x, N_y) + 8k,&  \:\: 4k < \min(N_x, N_y) \\
    N_x N_y, &\:\: 4k > \max(N_x, N_y) \\
    4k\min(N_x, N_y),&  \:\: \text{otherwise}.
    \end{cases}
\end{align*}

\subsection{Matrix Product State Preparation (MPS)}

We continue by studying circuits implementing quantum tensor networks, which provide broadly useful ansatze for representing various types of entangled quantum states. The simplest example of applying qubit reuse to a quantum tensor network circuit occurs in the context of matrix product state (MPS) preparation.

\begin{figure}[!ht]
    \centering
    \includegraphics[width=\columnwidth]{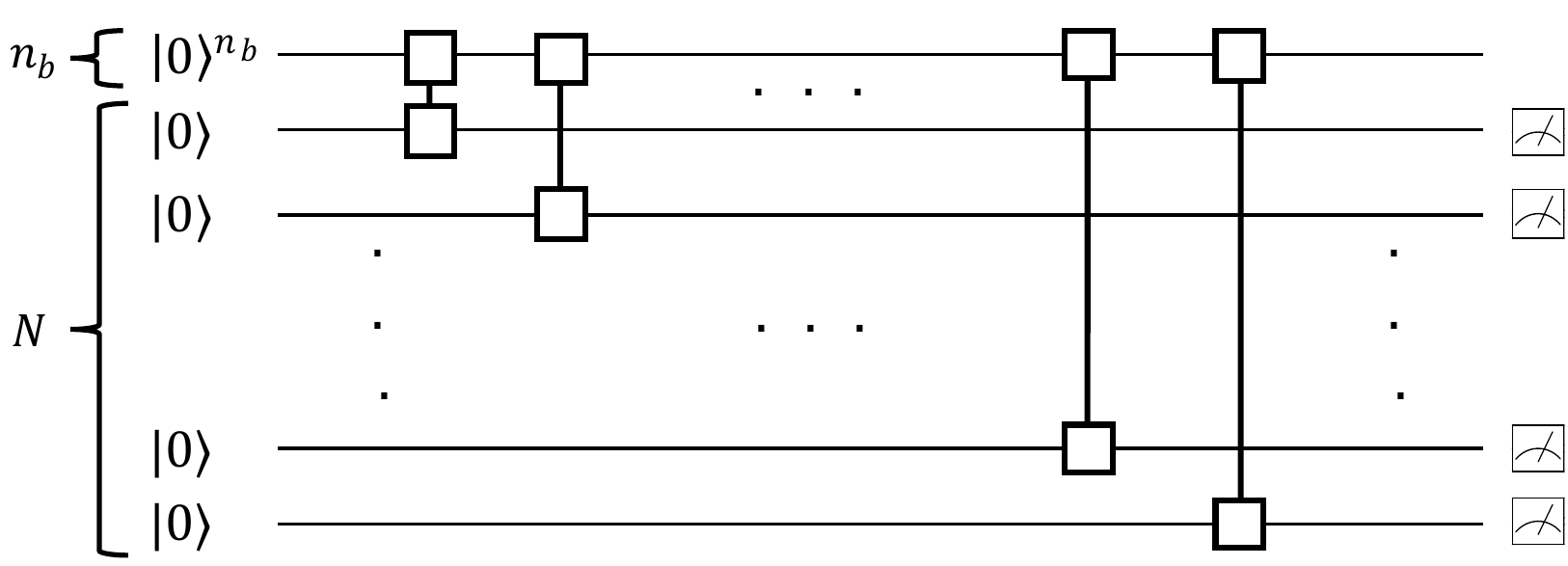}
    \caption{The MPS preparation circuit on $N$ qubits, using an MPS with $n_b = \lceil \log_2 \chi \rceil$ bond qubits, where $\chi$ is the bond dimension. Since the state preparation circuit consists of sequential gates applied to each of the physical qubits, the circuit can be rewritten using only $1 + n_b$ qubits by sequentially measuring and resetting each physical qubit.}
    \label{fig:MPS}
\end{figure}

As shown in Ref. \cite{PhysRevLett.95.110503}, by restructuring an N-site MPS into a suitable canonical form it can always be interpreted as a quantum circuit in which a register with Hilbert space dimension $\chi$ (representing the bond-space of the MPS) sequentially interacts with $N$ qubits (representing the physical legs of the MPS).  Therefore an MPS can always be constructed using $N+\lceil\log_2\chi\rceil$ qubits, as displayed in Fig.~\ref{fig:MPS}. As discussed in \cite{Huggins_2019,PhysRevResearch.1.023025, FossFeig2021}, owing to the sequential nature of the interactions between the ancilla register and the $N$ physical qubits, the full output of an MPS can be sampled with only $1+\lceil \log_2\chi\rceil$ qubits by resetting and reusing a single physical qubit after each local measurement.

\subsection{Tree Tensor Networks (TTN)}

\begin{figure}[!ht]
    \centering
    \includegraphics[width=\columnwidth]{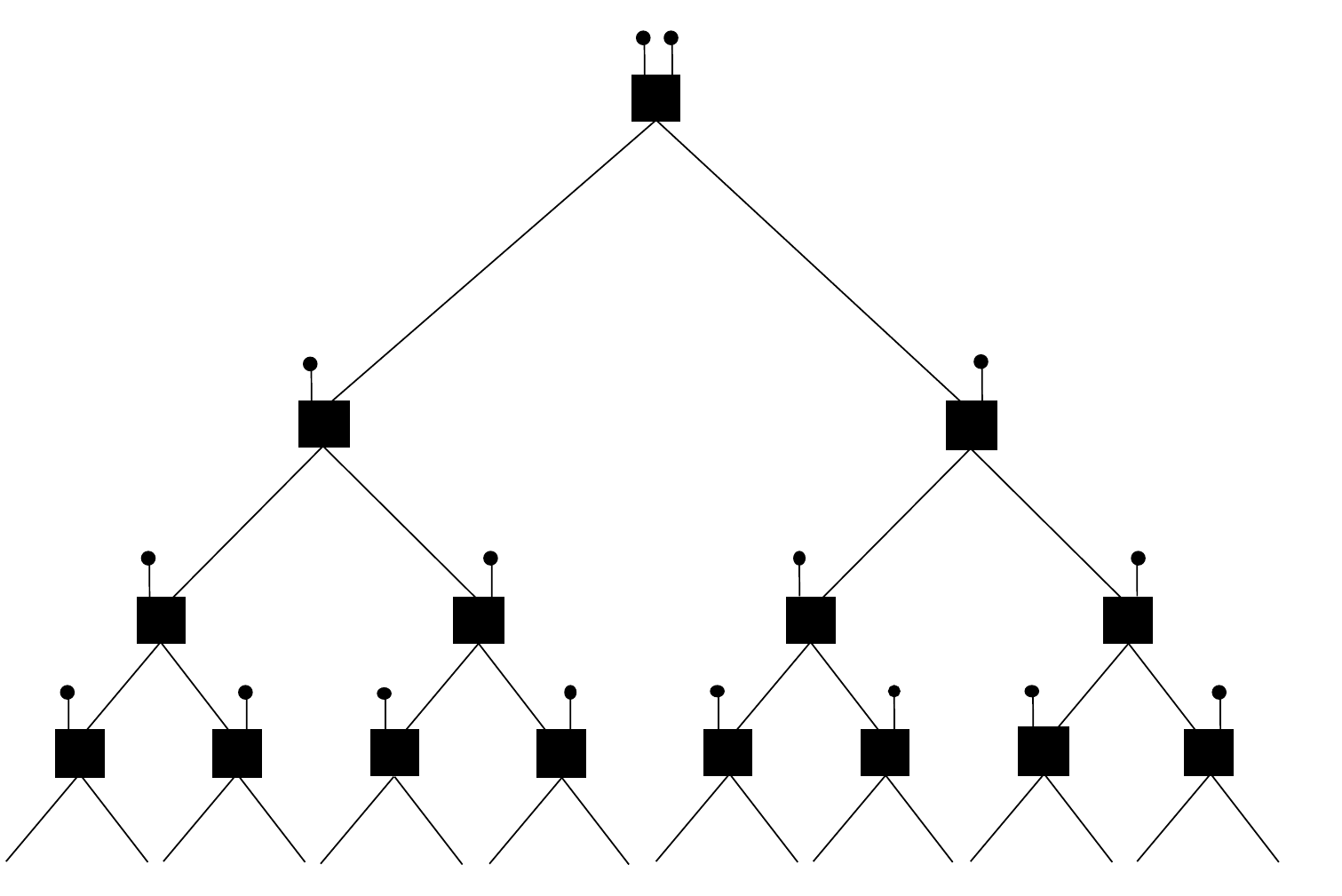}
    \caption{A depth-$4$ tree tensor network (TTN) generating a state on $16$ output qubits, read top to bottom. Each of the two legs of the topmost tensor connects to a depth-$3$ TTN as referenced in the text.}
    \label{fig:TTN}
\end{figure}

A more complex generalization of the one-dimensional entanglement generated by an MPS is provided by the tree tensor network (TTN), defined by placing isometries on the nodes of a binary tree, shown in Fig.~\ref{fig:TTN}. A depth-$D$ binary tree tensor network (TTN) generates a state of $2^D$ qubits, and can be written as a circuit by embedding the isometries into unitaries.  To determine how many qubits are required to sample the output of a general binary TTN, we will suppose that $N$ qubits are required to sample the output of a depth $D$ TTN and proceed inductively, as suggested in \cite{Huggins_2019}. A depth $D+1$ TTN is built by introducing a new top tensor, and attaching a depth $D$ TTN to each of its two outputs. Assuming for simplicity a bond-dimension of 2 (though this argument can extended to higher bond dimension), the output of the full depth $D+1$ TTN can be sampled by first inputting 2 qubits into the unitary representing the top tensor, after which point one qubit is idly waiting at the top of both depth $D$ TTNs sitting immediately below it. By assumption, the output of one depth $D$ TTN can be sampled with $N-1$ more qubits, since one qubit is already provided by the output of the top tensor. After sampling one of the two depth $D$ TTNs, the same $N-1$ qubits can be reused to sample the output of the second. Thus the full depth $D+1$ TTN can be sampled using $2+N-1=N+1$ qubits. Since a $D=1$ TTN can be sampled using 2 qubits, by induction it follows that the output of a depth $D$ TTN can be sampled using $D+1$ qubits.

Any TTN circuit can also be viewed as a coarse-graining procedure that converts an $N$-qubit input into a small number of qubits at the top of the network; in this mode of operation TTNs form natural classifiers \cite{Huggins_2019}.
For classical (product state) input data, the TTN classifier circuit is dual to the TTN generator circuit, and thus can also be executed using $D+1$ qubits.

\subsection{Multiscale Entanglement Renormalization Ansatz (MERA)}

\begin{figure*}
    \centering
    \includegraphics[width=0.98\textwidth]{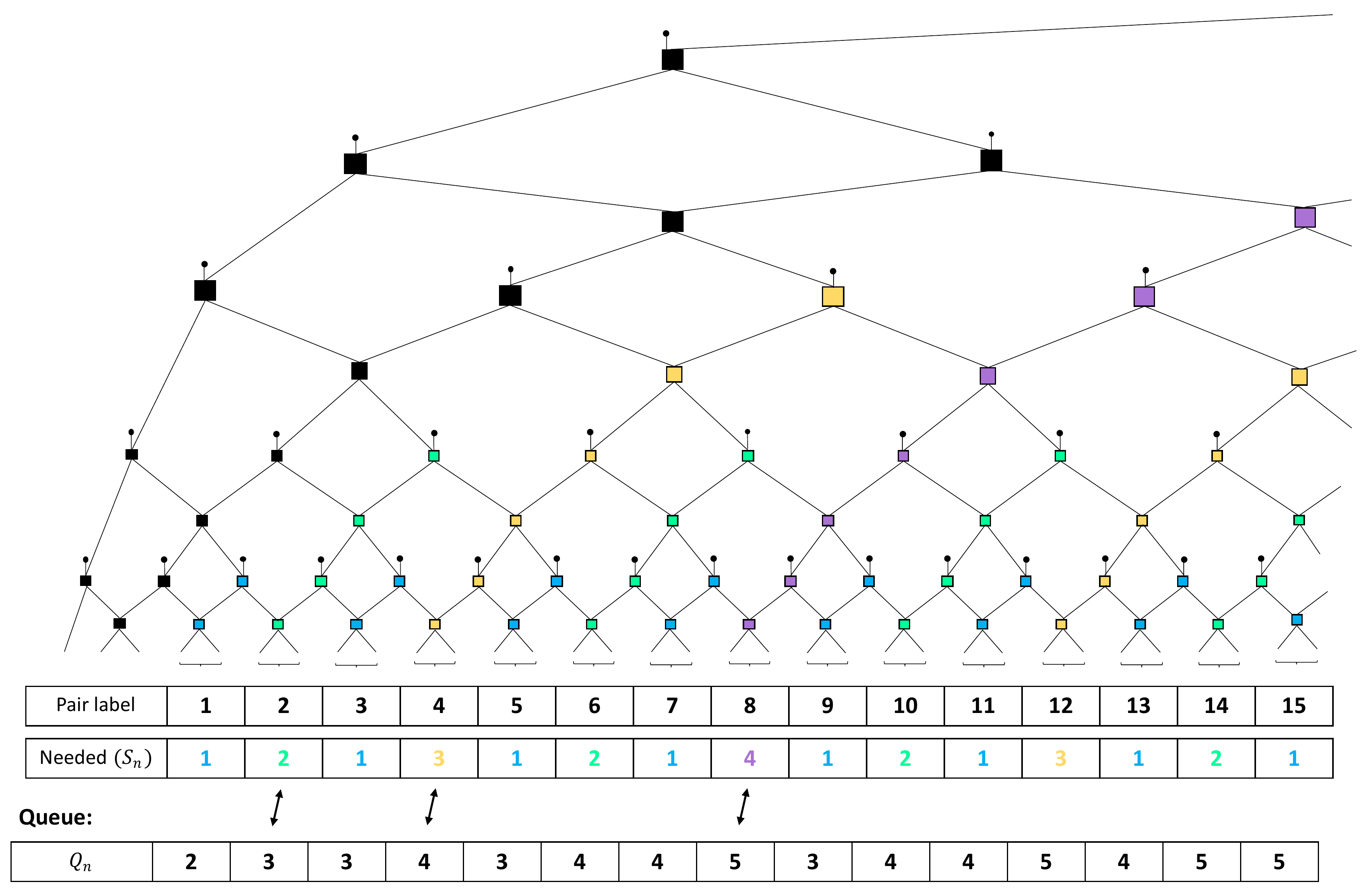}
    \caption{A section of an open boundary MERA circuit, with time going from top to bottom. Dots correspond to qubits initialized to $\ket{0}$ and squares to unitary gates (gates with an incoming $\ket{0}$ are isometries, while the others are disentanglers). Different colors of gates correspond to different causal cones of the output qubits at the bottom of the diagram. Output qubits come in pairs, which we label from left to right, with the fourth and fifth qubits from the left corresponding to pair $P_1$. The $S_n$ values indicate how many new isometries (and therefore new $\ket{0}$ qubits) need to be added to produce the causal cone of the qubits in pair $P_n$. The $Q_n$ values specify how many qubits are available for re-use from previously prepared causal cones.}
    \label{fig:MERA}
\end{figure*}

The multiscale entanglement renormalization ansatz (MERA) extends the tree-like architecture of a TTN by adding unitary disentanglers between neighboring branches of the tree.  Read from top to bottom as shown in Fig.~\ref{fig:MERA}, and interpreting the isometries as unitaries acting on additional initialized input qubits, one can view a MERA as a circuit that takes in $N$ qubits initialized in $\ket{0}$ and outputs the $N$-qubit state at the bottom of the MERA.  It is known to be a compact and efficiently contractible TN ansatz for critical states of matter \cite{PhysRevLett.101.110501}.

For a MERA with open boundary conditions the greedy measurement order is optimal for compressing the circuit.  To see this, begin by executing the past causal cone of the left-most qubit, which requires only $D+1$ input qubits and returns one to the ``reuse queue" of qubits that have already been measured and are available for reset and reuse.  The next step in a greedy approach is to implement the non-evaluated portion of the causal cone of the next two qubits from the left, which requires only $D-1$ qubits, one of which is borrowed from the reuse queue.  In this way, the first three outputs can be measured using $2D-1$ qubits, leaving 2 qubits in the reuse queue. It can be checked that any other choice of the first measured qubit would have required at least $2D-1$ qubits. Sequential qubits will be measured in pairs going down the line, which we index starting with the fourth and fifth qubit being $P_1$, and we denote the queue size at prior to measuring $P_1$ by $Q_1$. If one proceeds greedily from left-to right evaluating the causal cones of pairs $P_{n=2,3,...}$, the minimal required set of isometries for each pair is shown in Fig.~\ref{fig:MERA} using color coding. We denote the size of the set of isometries required to measure $P_n$ by $S_n$, which sets the number of qubits required to measure that pair.  The sequence $S$ is known as the ruler function \cite{rulerfunc}, and each value $S_n$ can be interpreted as the positions of the most significant bit (reading the least significant as the first) that gets flipped when incrementing from the binary representation of $n-1$ to that of $n$. Denote by $\Delta_n$ the number of qubits added to the reuse queue by measuring all pairs up to the one corresponding to the first instance of $n+1$ in the sequence $S$.  For example, $\Delta_1$ is obtained by executing the causal cone of pair $P_1$, which requires $1$ qubit, and then returning $P_1$ to the reuse queue, such that $\Delta_1=-1+2=1$. One can show that $\Delta_n$ obeys the recursion relation $\Delta_n=2(\Delta_n+1)-n$, which is satisfied (with proper boundary condition $\Delta_1=1$) by the solution $\Delta_n=n$.  Given that $Q_1=2$ qubits are in the queue prior to the evaluation of $P_1$'s causal cone, this ensures that there will always be $n+1$ qubits in the reuse queue at the first occurrence of $n$ in $S$ (e.g. $Q_{2^{j-1}}=j+1$, black arrows in the figure), such that we always have enough qubits in the reuse queue to execute the next causal cone without adding new qubits.  As we approach the right boundary of the MERA there will be times when fewer qubits are needed than suggested by this argument, but that only works in our favor. Thus we can sample the full output with only the $2D-1$ qubits required to measure the first three qubits.

\subsection{Quantum Convolutional Neural Networks (QCNN)}

By embedding each isometry into a unitary and either post-selecting on or discarding the padded outputs, any MERA can be read from bottom to top as a circuit that coarse-grains an $N$ qubit state, outputting a small number of qubits at the top.  This view lends itself naturally to the problem of classifying quantum data, and forms a natural quantum generalization of the convolutional neural network \cite{qcnn}. Note that the QCNN, as originally defined, involved classical feed-forward on measurements of padded qubits exiting the isometries; this does not change the causal structure of the MERA, and can be ignored for our purposes.  More generally, the conditional probabilities for measurement outcomes at the top of the QCNN do not change if we replace the measurement conditioned gates with quantum controlled gates, compile those gates into the unitary embedding of the isometry, and trace out the control qubit(s).  Assuming product-state inputs, this construction of a QCNN is exactly the dual circuit of the corresponding MERA, and thus can also be executed using $2D-1$ qubits in the worst case. However, we point out that the practical use of the QCNN to classify quantum data typically assumes the initial existence of a global many-body quantum state defined on $N = 2^D$ qubits rather than product-state inputs; in this case, the extent of qubit compression that can be accomplished will depend on the entanglement structure of the input state.

\subsection{Bernstein-Vazirani Algorithm}

\begin{figure}[!t]
    \centering
    \includegraphics[width=\columnwidth]{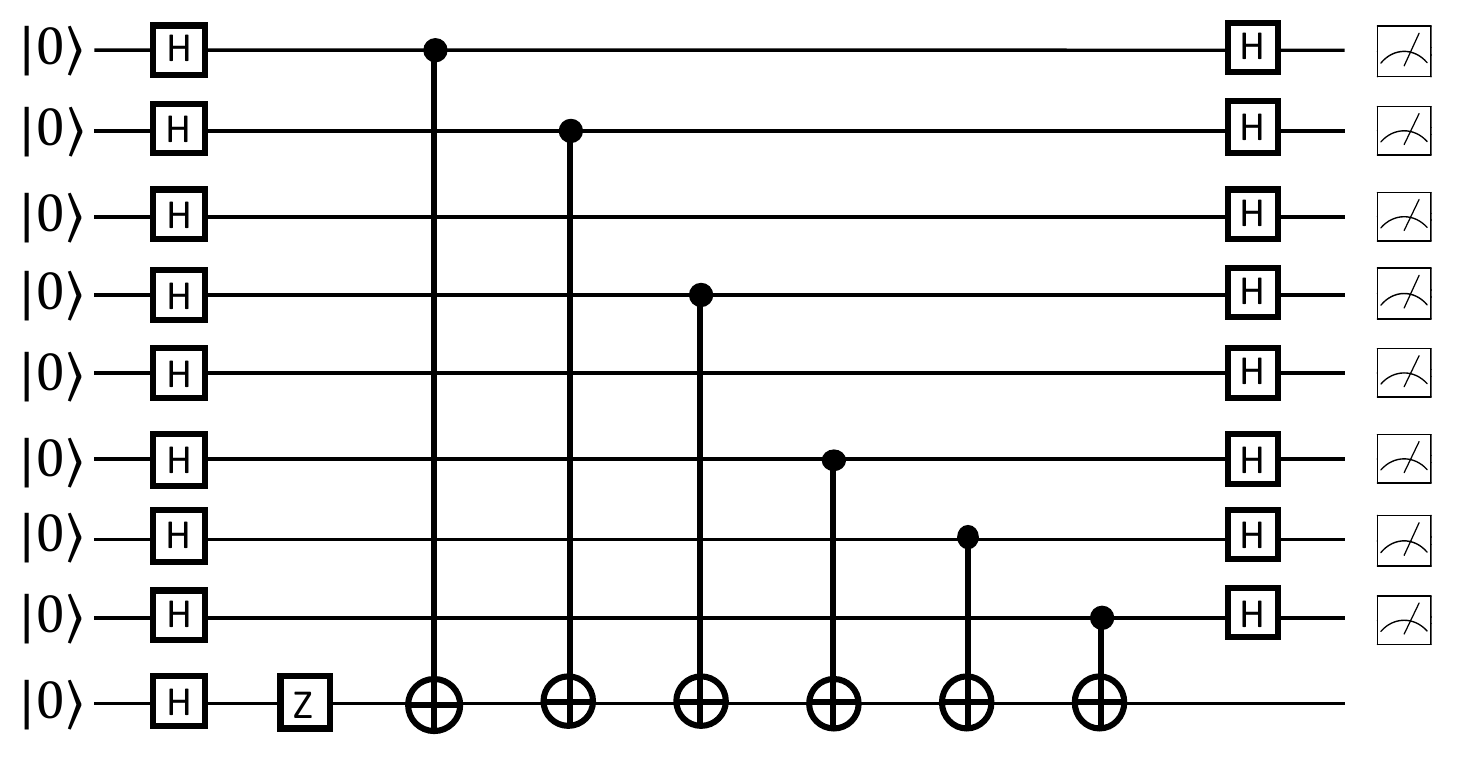}
    \caption{The Bernstein-Vazirani circuit on eight qubits, with the unitary oracle explicitly decomposed for the example of the hidden bitstring 11101011. Note that following the standard convention for encoding measurement bitstrings, the bottom-most qubit represents the left-most bit in the string.}
    \label{fig:bv_circ}
\end{figure}

The Bernstein-Vazirani algorithm solves the following problem: given a function $f: \{0,1\}^{N} \to \{0,1\}$ which is defined by a \emph{hidden bitstring} $s$ by $f(x) = s\cdot x \: (\textrm{mod } 2)$, find the bitstring $s$. The algorithm relies on the ability to query an oracle for the function $f$, so that classically $N$ function calls (on the $N$ unique bitstrings which are zero everywhere except for one register) are required to learn $f$. In the quantum case, the physical implementation of the unitary operator encoding the oracle typically requires knowledge of the hidden bitstring in advance, in which case the BV algorithm implemented as such is not a practical demonstration of quantum supremacy but rather a theoretical demonstration that given access to a true oracle only one query suffices in the quantum case (see Fig.~\ref{fig:bv_circ}). 

It is well-known \cite{H1BlogPost,BV2QubitBlogPost} that the BV algorithm can be implemented using only two qubits using mid-circuit measurement and reuse. In this section we simply state how this known result follows from the causal cone framework presented in this paper. Namely, none of the register qubits interact in the BV algorithm in the typical realization of the unitary oracle. Consequently, the causal cone of every register qubit contains input qubits that consist only of itself and of the shared ancilla qubit. Consequently, the BV algorithm defined on any number of qubits can always be executed using only two physical qubits. One physical qubit is used as the ancilla qubit and persists for the duration of the computation, while the other physical qubit represents a single register qubit. After the unitary oracle function is applied to the two qubits, the register qubit can subsequently be measured and reused as the next register until all registers are exhausted.

\vspace{.4cm}
The analytic results of this section are summarized in Table~\ref{tab:analytic_res}. The general effect of qubit-reuse compilation can be described as a sort of generalized holographic quantum dynamics for local physics and tensor network circuits, in that the scaling of the compiled qubit number is parametrically less than that of the original circuit.

\begin{table}[!ht]
$$\begin{tabular}{ccc}\\
Algorithm & Original Qubit \#\:\: & Compiled Qubit \# \\ \hline
$k$-layer brickwork 1D & $N$ & $4k$ \\
$k$-layer brickwork 2D & $N^2$ & $(4k-2)N + 8k$  \\
MPS, bond dim. $\chi$ \cite{FossFeig2021} & $N + \lceil\log_2 \chi \rceil$  &  $1 + \lceil\log_2 \chi \rceil$\\
depth--$D$ binary TTN & $2^D$ & $D+1$ \\
depth--$D$ binary MERA & $2^D$ & $2D-1$ \\
depth--$D$ binary QCNN & $2^D$ & $2D-1$ \\
Bernstein-Vazirani & $N$ & $2$
\end{tabular}$$
\caption{The minimum number of qubits required to execute certain structured quantum circuits after qubit-reuse compilation. For the $k$-layer brickwork circuits we assume that $4k$ is smaller than all dimensions. \label{tab:analytic_res}}
\end{table}

\section{Numerical Experiments and Benchmarking} \label{sec:numerics}

We expect many practical examples of circuits on which qubit reuse may be helpful to lie outside of the analytically tractable categories discussed in the previous section. To this end, in this section we numerically study the amount of compression that can be achieved in the quantum approximation optimization algorithm (QAOA). We study the algorithmic time complexity (runtime) and solution quality (compiled qubit number) for QAOA MaxCut circuits at various depths on random unweighted three-regular (U3R) graphs. MaxCut on random U3R graphs is a standard benchmarking case for QAOA on near-term quantum devices (see \cite{google1, google2, rydberg_qaoa} for some recent investigations of QAOA on hardware), since the number of gates matches the number of edges in the graph and scales only linearly with the number of vertices. In general, MaxCut QAOA is an ideal near-term algorithm for qubit reuse as it is a shallow, wide circuit with fairly sparse gate connectivity, so the typical size of causal cones can be expected to be small relative to the size of the original set of qubits.

The QAOA unitary \cite{original_qaoa} takes the form of alternating applications of a mixing unitary $U_B(\beta_n) = e^{-i\beta_n H_B}$ and a phase-splitting cost unitary $U_C (\gamma_n) = e^{-i\gamma_n H_C}$,
\begin{align}
    U(\vec{\beta}, \vec{\gamma}) = \prod_{n=1}^p U_B(\beta_n) U_C (\gamma_n)
\end{align}
where $H_B = \sum_i X_i$ and $H_C$ encodes the cost Hamiltonian of the combinatorial problem. The unitary depends on $2p$ parameters $\beta_1, \ldots, \beta_p$ and $\gamma_1, \ldots, \gamma_p$ and the product is conventionally ordered so that the terms with $\beta_1$ and $\gamma_1$ are applied first. For the MaxCut problem, the cost Hamiltonian $H_C$ has a standard encoding as a quadratic unconstrained binary optimization (QUBO) Hamiltonian, taking the form
\begin{align}
    H_C = \frac12 \sum_{(i,j) \in E} w_{ij} (1-Z_i Z_j)
\end{align}
where the coefficient $w_{ij}$ is the weight of edge $(i,j)$. Since we take the graph to be unweighted in all cases, all coefficients $w_{ij} = 1$.

The QAOA protocol finds the parameters $\vec{\beta}, \vec{\gamma}$ by variational minimization. A classical optimization algorithm is used to search for parameters $\vec{\beta}^{\ast}, \vec{\gamma}^{\ast}$ that minimize
\begin{align}
    \langle H_C \rangle = \langle \psi_0 | U(\vec{\beta}, \vec{\gamma})^{\dagger} H_C U(\vec{\beta}, \vec{\gamma}) | \psi_0\rangle \label{eq:exp_val}
\end{align}
with $|\psi_0\rangle = |+\rangle^{\otimes N}$ by convention. For the purposes of studying qubit reuse, the circuits we consider evaluate $U(\vec{\beta}, \vec{\gamma})|+\rangle^{\otimes N}$ and measure all qubits at the end. 

Throughout this section, we will primarily be concerned with comparing three different algorithms for compiling circuits with qubit reuse: the local greedy algorithm, the local greedy algorithm with an additional brute force search over the first measured qubit, and the exact CP-SAT solution where it is viable. Depending on the circuit structure and the amount of compression that is possible, it becomes impractical to execute the CP-SAT model at around $N \sim 30-50$ qubits in the original circuit, so we will also compare only the heuristic algorithms at significantly larger $N$. The time benchmarking presented in this section was evaluated on a small personal laptop, effectively as a model of the realistic computation time that can be expected by a typical user desiring to use qubit-reuse compilation for algorithm development; it is possible that high-performance computing techniques can slightly extend the regime in which the algorithms presented herein are feasible.

\begin{figure}[!t]
    \centering
    \includegraphics[width = .5\textwidth]{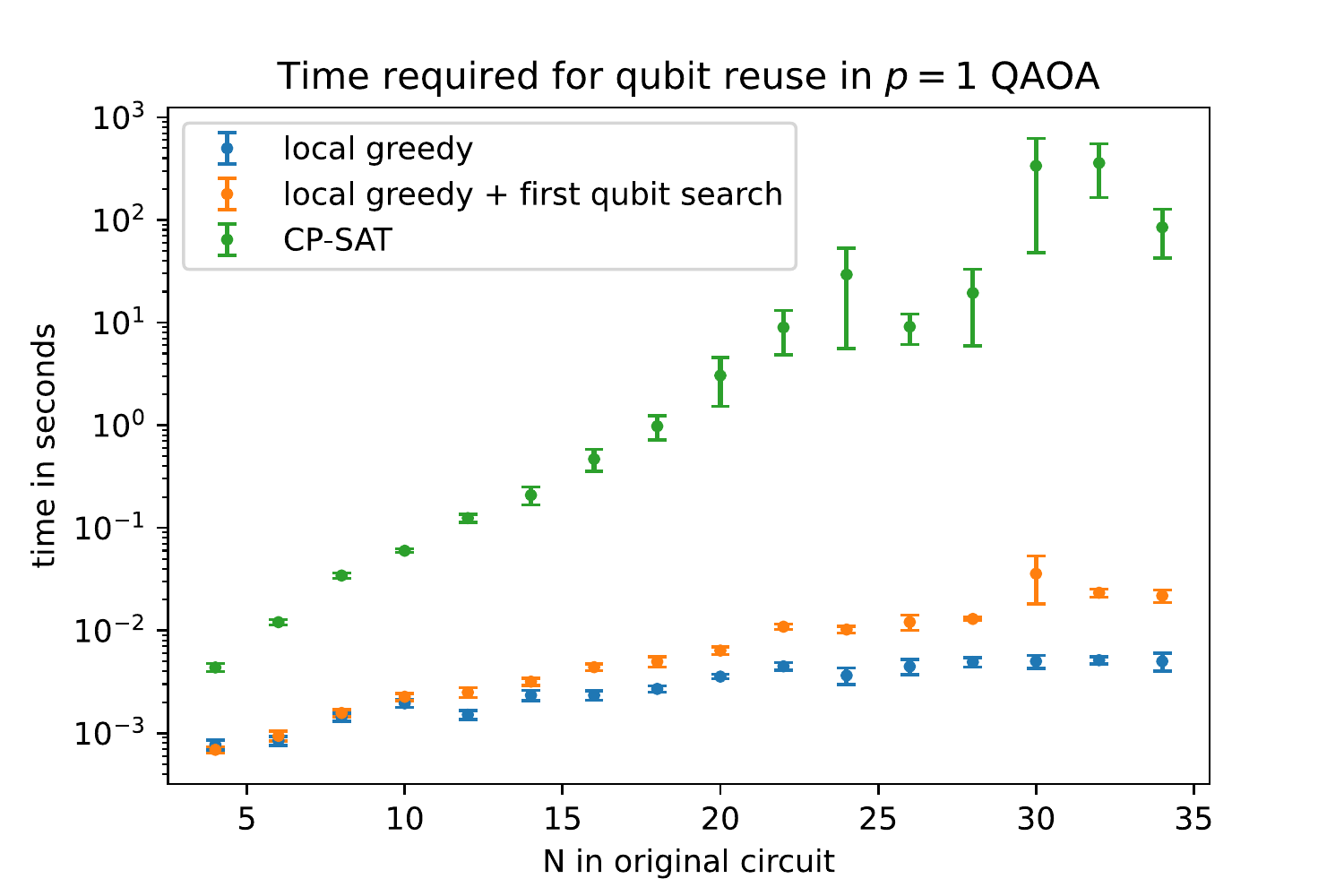}
    \caption{The time required in seconds to execute each qubit-reuse algorithm as a function of the number of qubits in the original circuit, averaged over $100$ instances of random U3R graph MaxCut $p=1$ QAOA circuits. The plotted uncertainties correspond to the error on the mean in this and all below plots.}
    \label{fig:exp_results1}
\end{figure}

\begin{figure}[!t]
    \centering
    \includegraphics[width = .5\textwidth]{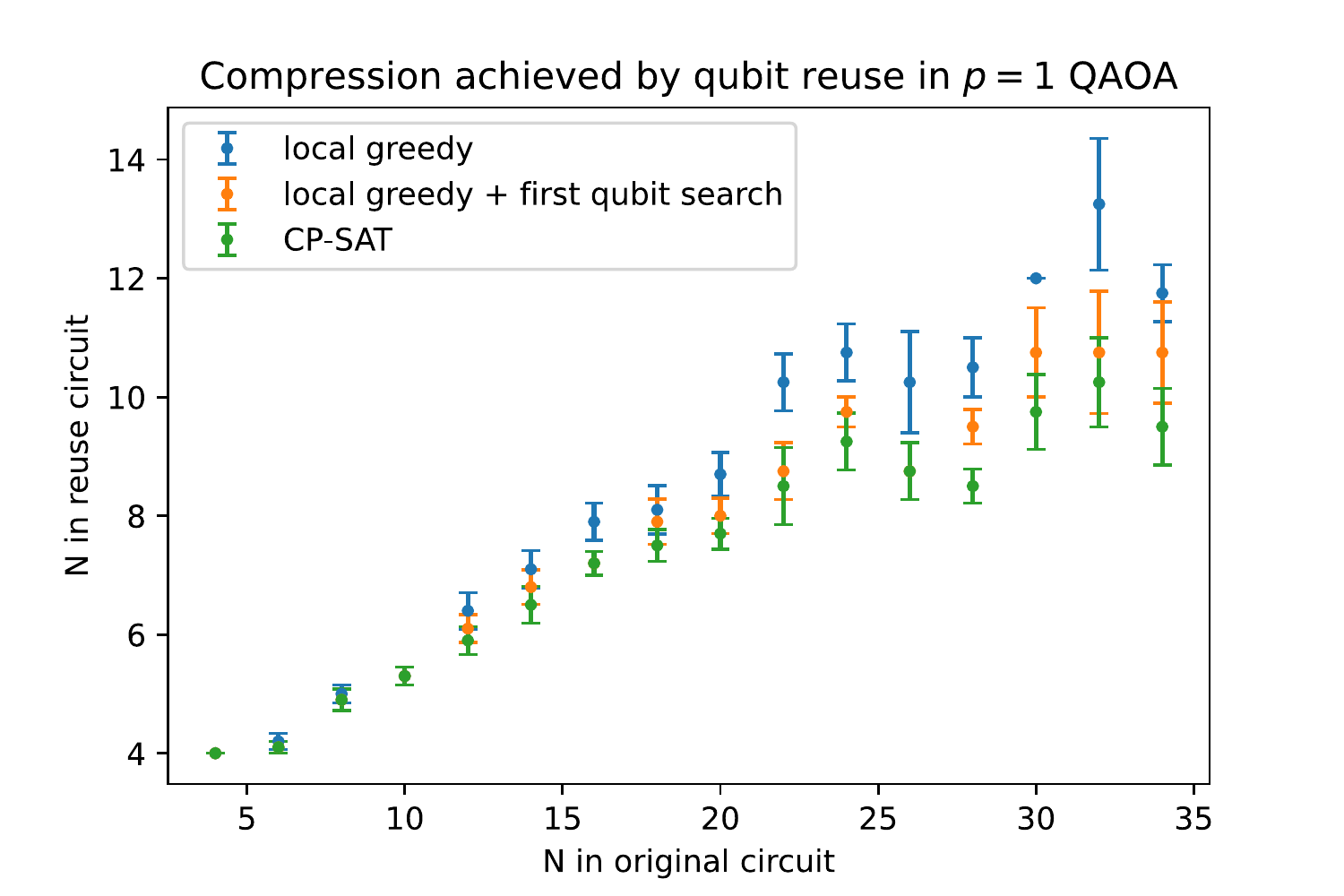}
    \caption{The compiled qubit number as a function of the number of qubits in the original circuit, averaged over $100$ instances of random U3R graph MaxCut $p=1$ QAOA circuits.}
    \label{fig:exp_results2}
\end{figure}

\begin{figure}[!t]
    \centering
    \includegraphics[width = .5\textwidth]{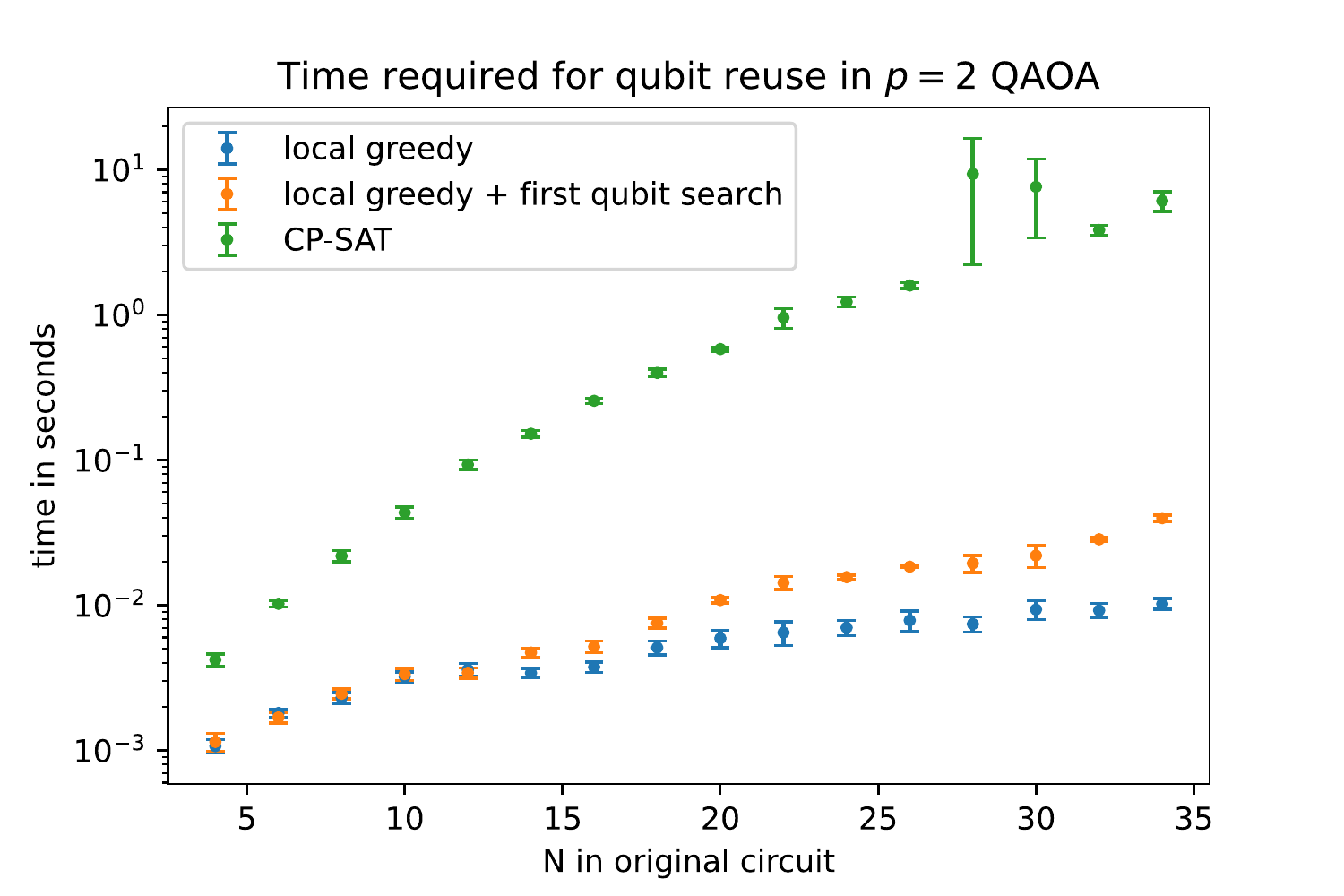}
    \caption{The time required in seconds to execute each qubit reuse algorithm as a function of the number of qubits in the original circuit, averaged over $100$ instances of random U3R graph MaxCut $p=2$ QAOA circuits.}
    \label{fig:exp_results3}
\end{figure}

\begin{figure}[!t]
    \centering
    \includegraphics[width = .5\textwidth]{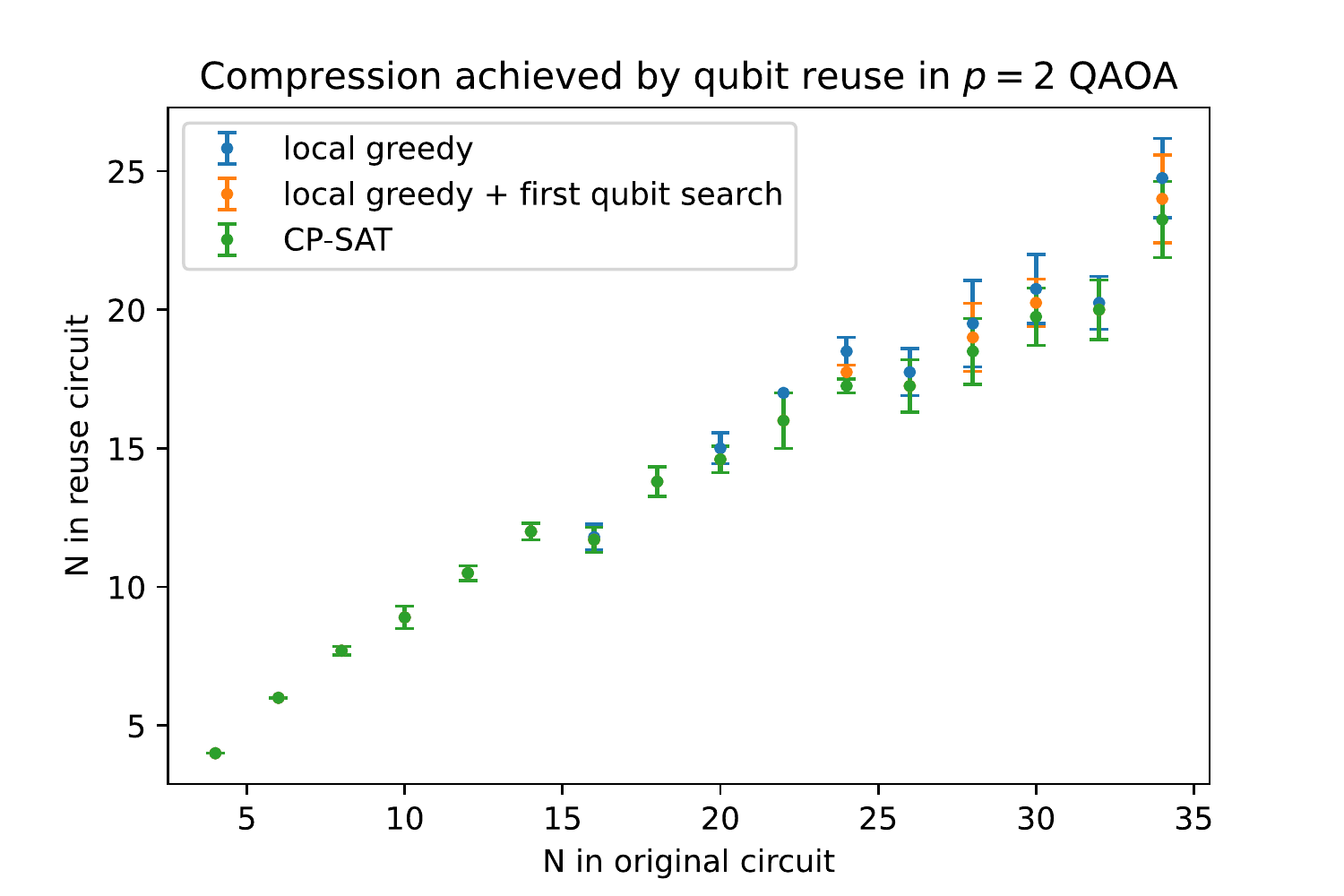}
    \caption{The compiled qubit number as a function of the number of qubits in the original circuit, averaged over $100$ instances of random U3R graph MaxCut $p=2$ QAOA circuits.}
    \label{fig:exp_results4}
\end{figure}

We begin by examining the performance of the three algorithms for $p = 1$ (Fig.~\ref{fig:exp_results1}, Fig.~\ref{fig:exp_results2}) and $p = 2$ (Fig.~\ref{fig:exp_results3}, Fig.~\ref{fig:exp_results4}) MaxCut QAOA. For each fixed qubit number, we evaluate the qubit reuse algorithms on $100$ random U3R graphs generated using the NetworkX package \cite{networkx}.

As expected, it is clear from Fig.~\ref{fig:exp_results1} that the CP-SAT model is superpolynomially scaling and rapidly becomes impractical for circuits of more than a few dozen qubits. Furthermore, from Fig.~\ref{fig:exp_results2} it is apparent that the performance of the heuristics is only slightly worse than optimal at these small qubit numbers, especially when the additional brute force search is employed on the first qubit measured.

In Fig.~\ref{fig:exp_results3} and Fig.~\ref{fig:exp_results4} one can see that for $p = 2$, although the runtime of the heuristic algorithms slightly increases from $p = 1$, the CP-SAT model runs about an entire order of magnitude more quickly. This can be explained by the behavior of the classical optimization techniques underlying the CP-SAT solver, which iteratively rule out sections of the solution search space. Because the causal cones in $p = 2$ QAOA are larger than that of $p = 1$ QAOA, the CP-SAT model is more highly constrained (see \eqref{eq:second_constraint}) in the $p = 2$ case, allowing more of the search space to be ruled out faster. We further see that all three algorithms perform nearly comparably and achieve significantly less compression of the original circuit in the $p = 2$ case.

\begin{figure}[!t]
    \centering
    \includegraphics[width = .5\textwidth]{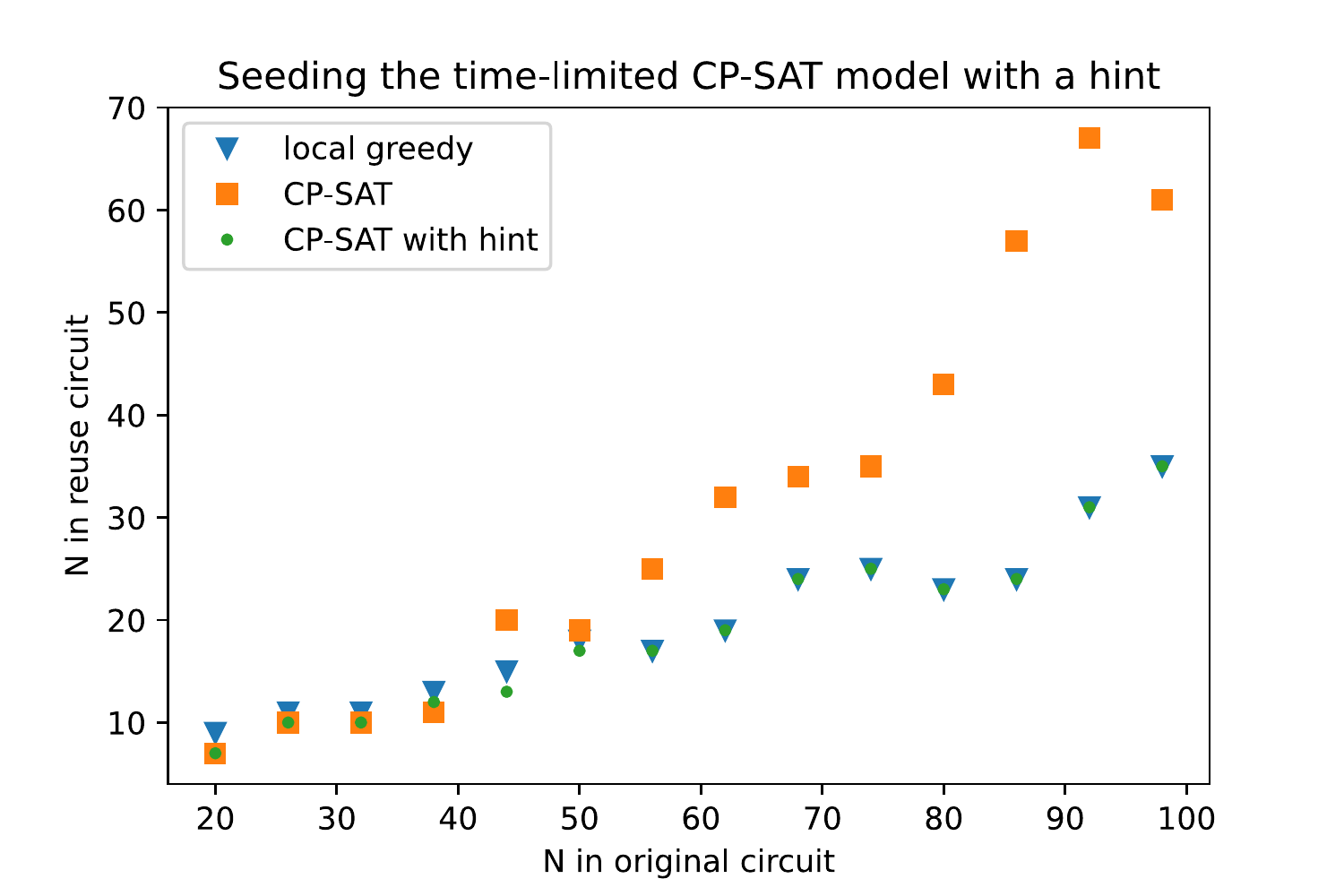}
    \caption{The compiled qubit number as a function of the number of qubits in the original circuit, for one instance of random U3R graph MaxCut $p = 1$ QAOA at each point.}
    \label{fig:exp_results9}
\end{figure}

The CP-SAT model also permits seeding with an initial ``hint" solution that satisfies the provided constraints, attempting to improve upon this solution. This suggests an obvious method to improve upon the heuristic solutions at larger $N$ by handing off the heuristic solution as input to the CP-SAT model. In Fig.~\ref{fig:exp_results9} we compare the amount of compression achieved in $p = 1$ MaxCut QAOA by the local greedy algorithm to that produced by the CP-SAT model both with and without the hint. To realistically bound the amount of time available for attempting qubit-reuse compilation, the CP-SAT model was limited to ten minutes to attempt to find a solution both with and without the hint.

Fig.~\ref{fig:exp_results9} displays several different regimes owing to the time limit placed on the CP-SAT model. At sufficiently small $N$, the CP-SAT model solves the qubit-reuse problem exactly in under ten minutes, so the hint provides no added value. Around $N \sim 40$ there is a transition, where ten minutes is no longer sufficient for CP-SAT to optimally solve the qubit-reuse problem. In this case, seeding with the hint may produce a better solution than either the heuristic or CP-SAT solutions without the hint, though the point at $N = 38$ shows that it is also possible to get stuck in a local optimum when starting from the hint that does not beat the unseeded model. Finally, at $N \gtrsim 60$, ten minutes is clearly insufficient for the CP-SAT model to find an optimal solution, since it is beaten by the local greedy algorithm, and it is also insufficient for the CP-SAT model to improve on the heuristic solution.

\begin{figure}[!t]
    \centering
    \includegraphics[width = .5\textwidth]{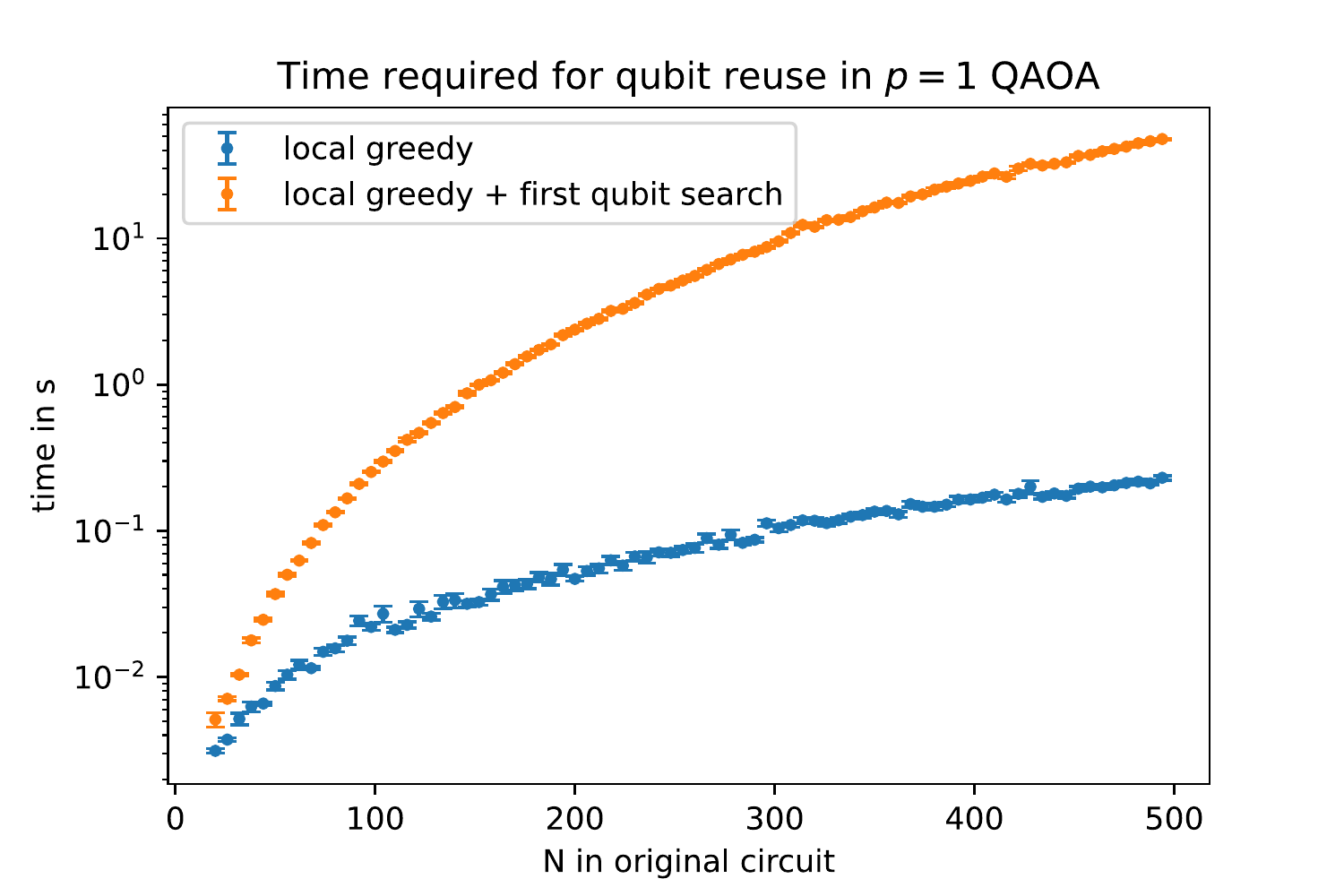}
    \caption{The time required in seconds to execute each qubit reuse algorithm as a function of the number of qubits in the original circuit for random U3R graph MaxCut $p = 1$ QAOA circuits. The data is averaged over $100$ instances for $N < 100$ and $20$ instances for $N > 100$.}
    \label{fig:exp_results5}
\end{figure}

\begin{figure}[!t]
    \centering
    \includegraphics[width = .5\textwidth]{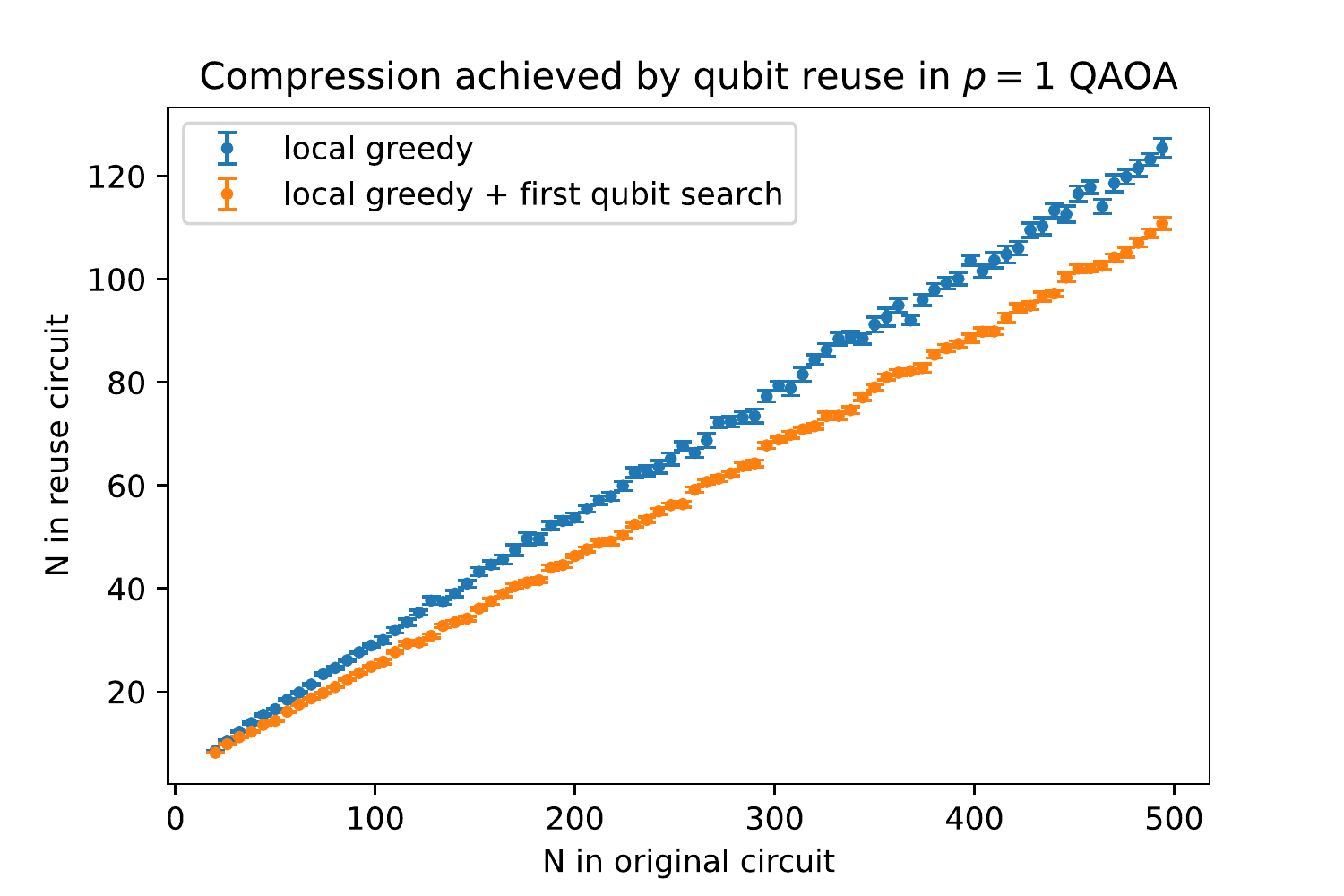}
    \caption{The compiled qubit number as a function of the number of qubits in the original circuit for random U3R graph MaxCut $p = 1$ QAOA circuits. The data is averaged over $100$ instances for $N < 100$ and $20$ instances for $N > 100$.}
    \label{fig:exp_results6}
\end{figure}

Looking ahead, it is of interest to determine how well qubit-reuse compilation will perform when algorithms are scaled to several hundred or several thousand qubits in size. To this end, we study the performance of the two heuristic algorithms outside of the regime in which the CP-SAT model is practical. The results (Fig.~\ref{fig:exp_results5} and Fig.~\ref{fig:exp_results6}) show that enhancing the local greedy algorithm with the brute force search is viable out to at least $N \sim 500$, where the qubit-reuse compilation takes about a minute for the improved local greedy algorithm, and fractions of a second for the local greedy algorithm without this improvement. Furthermore, the brute force search over the first qubit results in a $13\%$ reduction on average in the number of qubits required to execute the circuit compared to the greedy algorithm with no brute force search, requiring only about one-quarter of the original number of qubits on average. The upshot is that for near-term applications with dozens or at most hundreds of qubits, it is essentially always preferable to include this improvement when executing the local greedy algorithm.

\begin{figure}[!t]
    \centering
    \includegraphics[width = .5\textwidth]{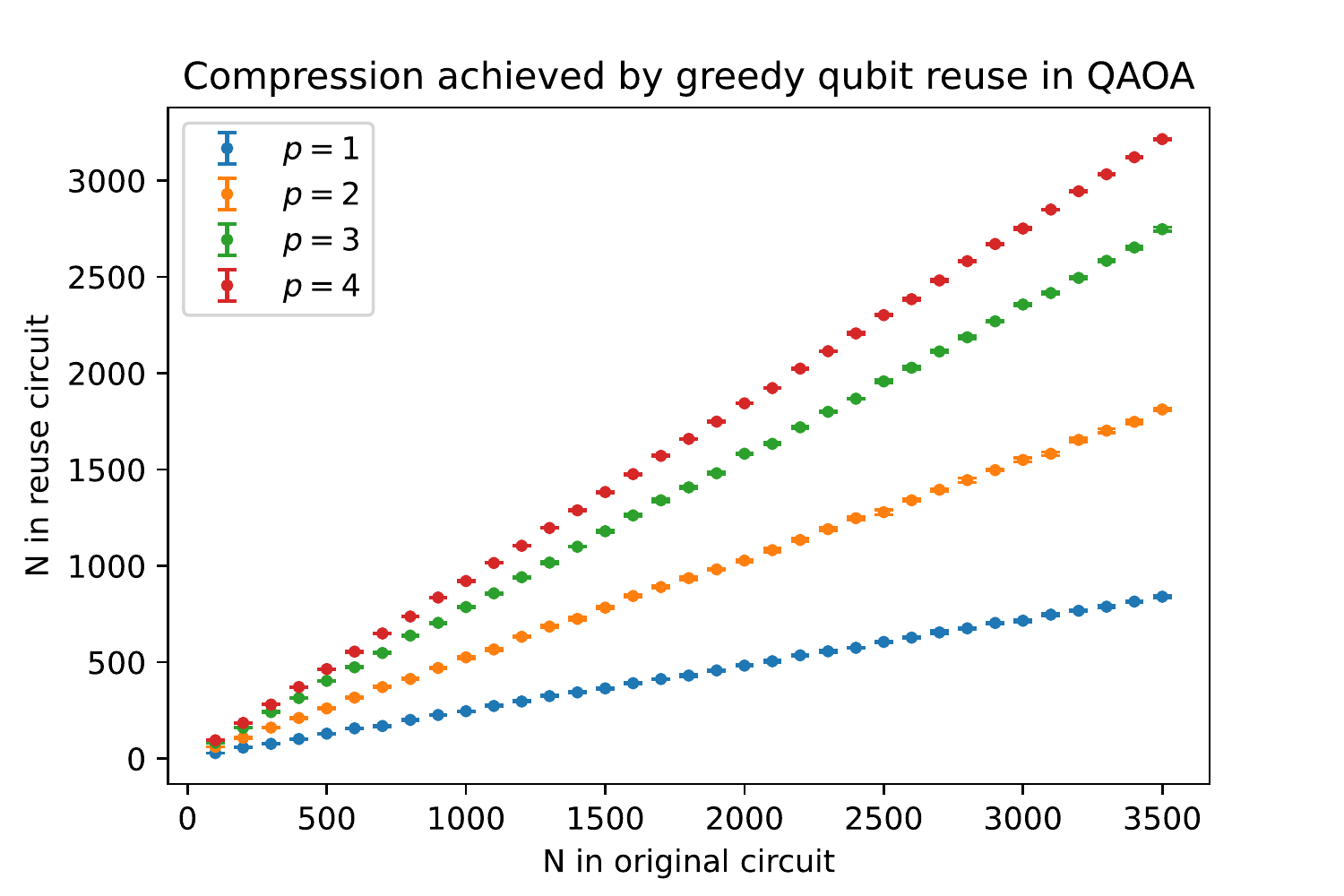}
    \caption{The compiled qubit number as a function of the number of qubits in the original circuit, averaged over $10$ instances of random U3R graph MaxCut QAOA circuits at different depths $p$.}
    \label{fig:exp_results7}
\end{figure}

\begin{figure}[!t]
    \centering
    \includegraphics[width = .5\textwidth]{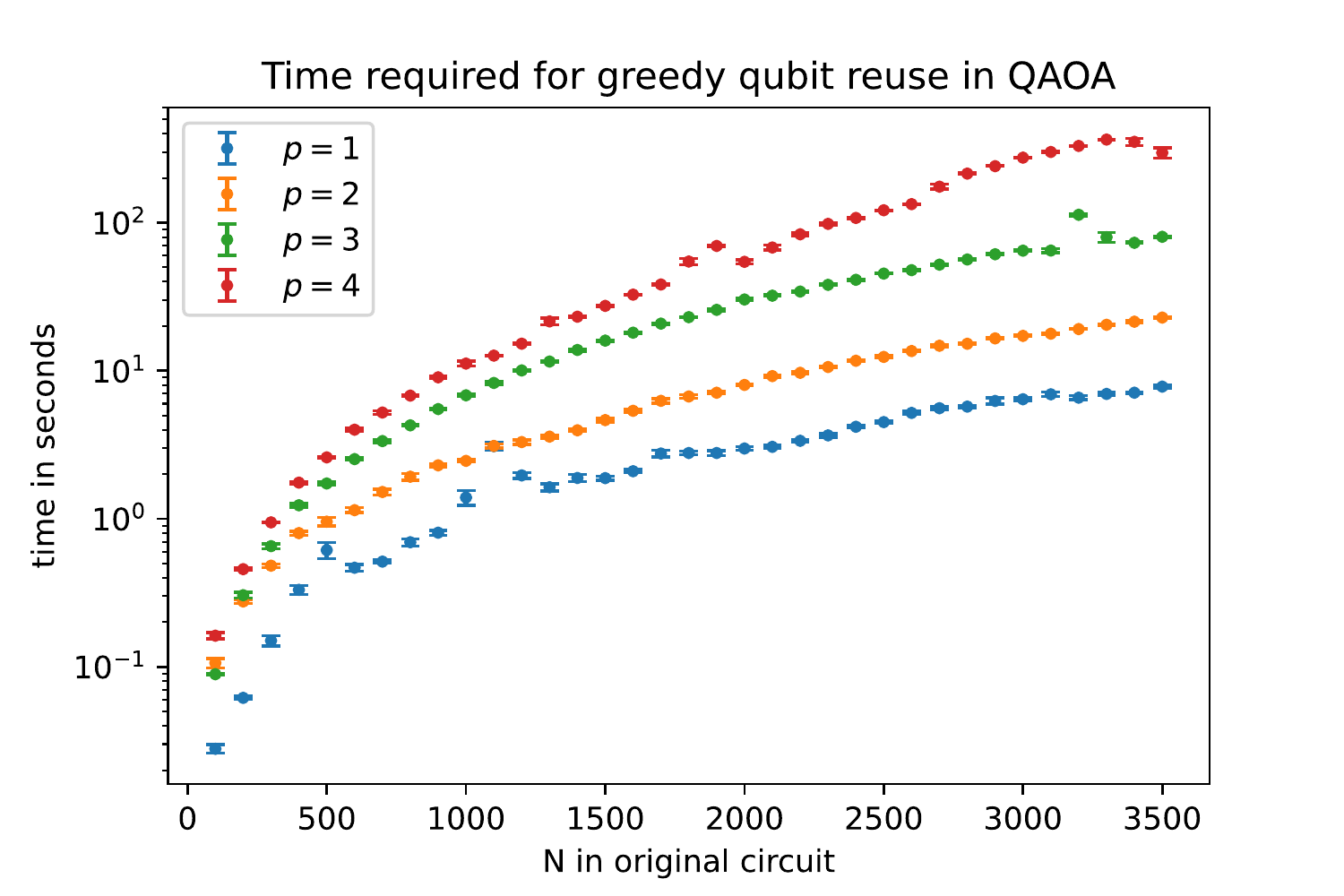}
    \caption{The time required in seconds to execute the local greedy algorithm as a function of the number of qubits in the original circuit, averaged over $10$ instances of random U3R graph MaxCut QAOA circuits at different depths $p$.}
    \label{fig:exp_results8}
\end{figure}

Finally, it is prudent to test the limits of the local greedy algorithm (with no brute force search improvement) both in terms of time expenditure and how well it scales with deeper circuits. To this end, in Fig.~\ref{fig:exp_results7} and Fig.~\ref{fig:exp_results8} we display the performance and efficiency of the local greedy algorithm out to several thousand qubits in the original circuit. The results show that the local greedy algorithm scales quite well, running efficiently in under a minute on circuits with several thousand qubits. The amount of time required also increases with the circuit depth (number of iterations $p$), although we expect both the time and amount of achievable compression to plateau at sufficiently large $p$, once the causal cones are the size of the entire set of qubits. In particular, we expect no compression of the circuit to be achievable at sufficiently large depth, and already by $p = 4$ there is very little savings in the overhead required to execute the circuit achieved by the compilation.

\section{Experimental Demonstration}
\label{sec:exp_demo}

We now experimentally demonstrate the practical functionality of qubit reuse by solving an eighty-qubit combinatorial problem with QAOA, using circuits executed on only twenty qubits. We study a specific U3R MaxCut instance by generating a random U3R graph on eighty vertices using the NetworkX Python package, for which the corresponding direct QUBO encoding requires eighty qubits. We expect qubit-reuse to be most effective in shallower circuits, so $p=1$ iteration of QAOA was chosen for this demonstration.
\begin{figure}[!t]
    \centering
    \includegraphics[width = .5\textwidth]{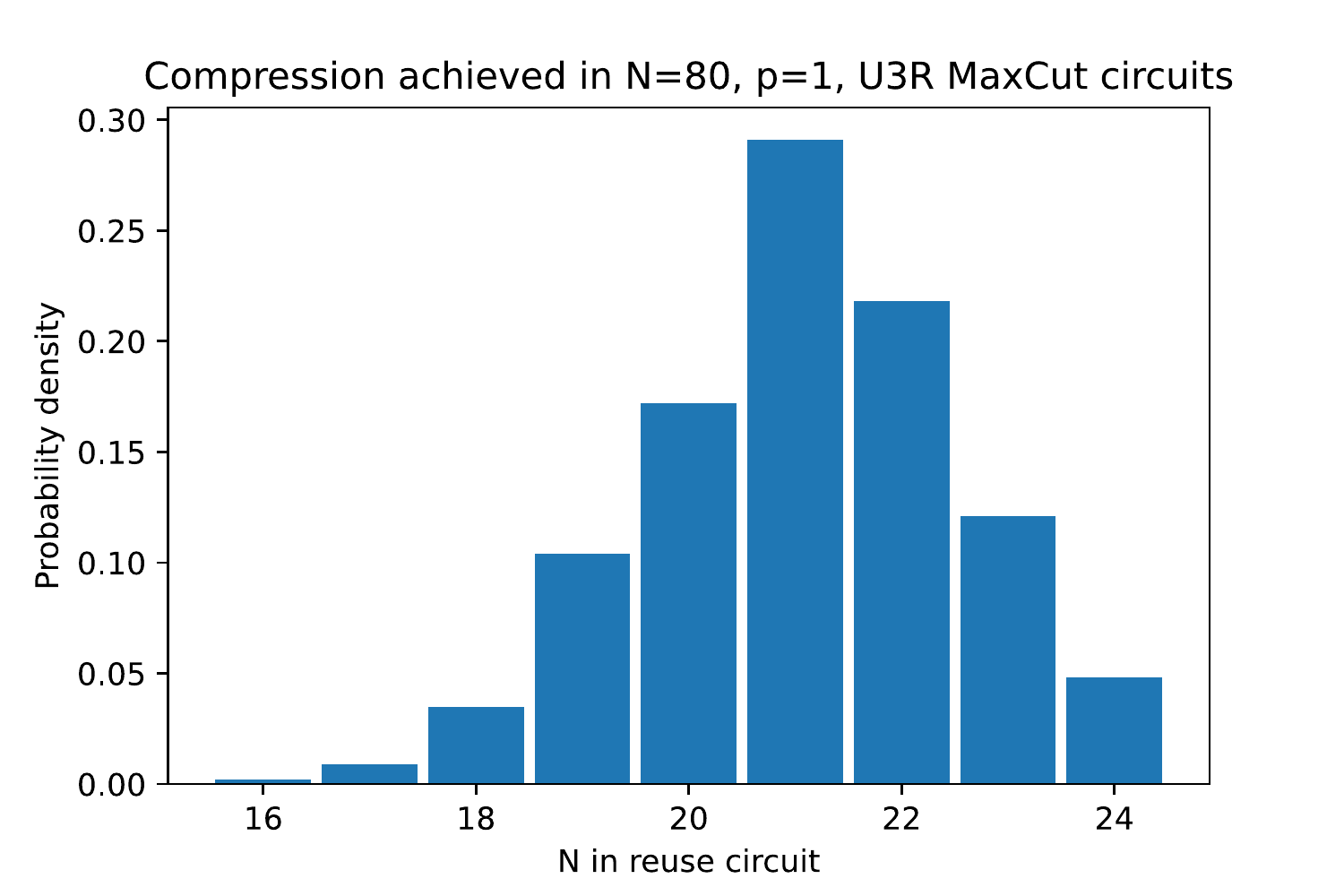}
    \caption{The number of qubits required to execute $N=80$, $p=1$ MaxCut circuits for 1000 instances of random U3R graphs, using the local greedy heuristic algorithm with brute force search on the first qubit.}
    \label{fig:compression}
\end{figure}
A histogram of the typical amount of compression that can be achieved for such eighty-qubit $p=1$ QAOA circuits is displayed in Figure~\ref{fig:compression} for 1000 instances of random U3R graphs, which demonstrates that the typical eighty-qubit U3R MaxCut instance can be expected to compress to a circuit with $21.1 \pm 1.5$ qubits on average. In particular, $32\%$ of instances compress to twenty qubits or less, so this amount of compression is a reasonable expectation for a typical instance. As a result, the first random U3R graph that was sampled allowed for compression to the Quantinuum H1-1 hardware limitation of twenty qubits. In order to compute the minimal number of qubits required to execute these circuits, we executed the local greedy algorithm with the additional improvement of brute force search over the first qubit as described above.

For this random U3R graph, we generated the $p=1$ QAOA circuit as a function of parameters $\beta_1$ and $\gamma_1$ that were initialized at a random point. Owing to the expense of numerically approximating derivatives with many function calls, we chose a derivative-free optimizer that performed the best comparatively in simulations, the BOBYQA optimizer (Bound Optimization BY Quadratic Approximation) \cite{BOBYQA} as implemented in the Py-BOBYQA package \cite{PyBOBYQA}. This optimizer computes interpolating points to generate a quadratic approximation to the objective function within a trust region of a particular size. We used the out-of-the-box settings for this optimizer, wherein the convergence criterion is taken to be the size of the trust region becoming smaller than $10^{-8}$.

Starting from the random initialization we chose, the full QAOA optimization procedure was run on the Quantinuum H1-1 trapped ion quantum computer \cite{H1-1}. Each circuit was run for 100 shots, for a total of roughly three minutes per circuit of machine time. This number of shots was chosen by simulation of the optimization procedure as sufficiently large enough to statistically measure decreases in the objective function while remaining small enough to make the machine time feasible. A total of seventy-eight circuits were run before the convergence criterion of the optimizer was met.

\begin{figure}[!t]
    \centering
    \includegraphics[width = .5\textwidth]{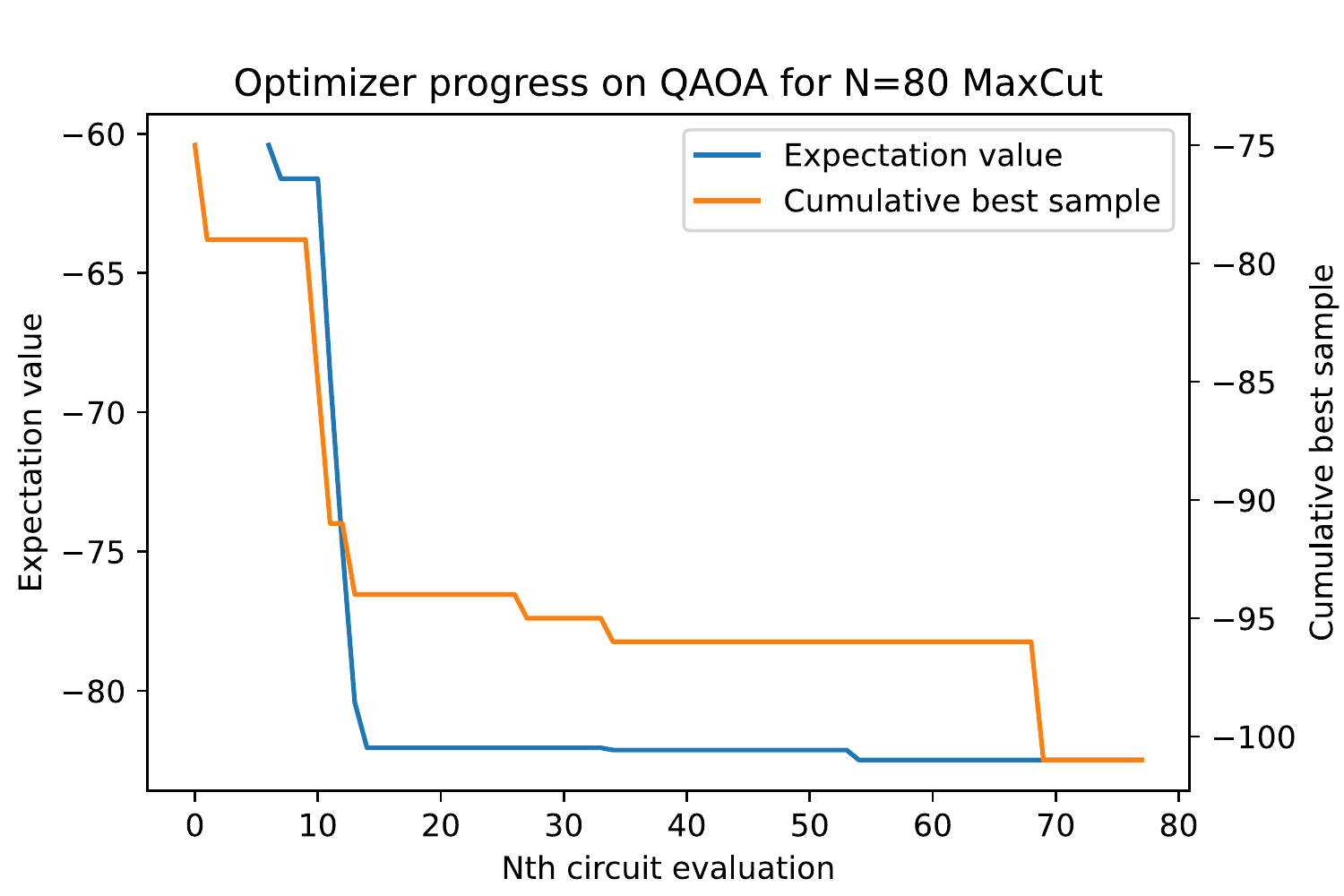}
    \caption{Progress of the objective function over the course of the BOBYQA optimization (blue), overlaid on the cumulative best sample for all circuits submitted so far (orange). The first six circuits are used to seed the quadratic approximation in the optimizer, which does not record progress before that point.}
    \label{fig:exp_results}
\end{figure}

In Fig.~\ref{fig:exp_results} we show the progress of the objective as evaluated at the best point selected by the optimizer so far. As in \eqref{eq:exp_val}, the expectation value is used as the objective function for the optimizer, but we also plot the progress of the best sampled cut value among all of the shots taken.

\begin{figure}[!t]
    \centering
    \includegraphics[width=0.5\textwidth]{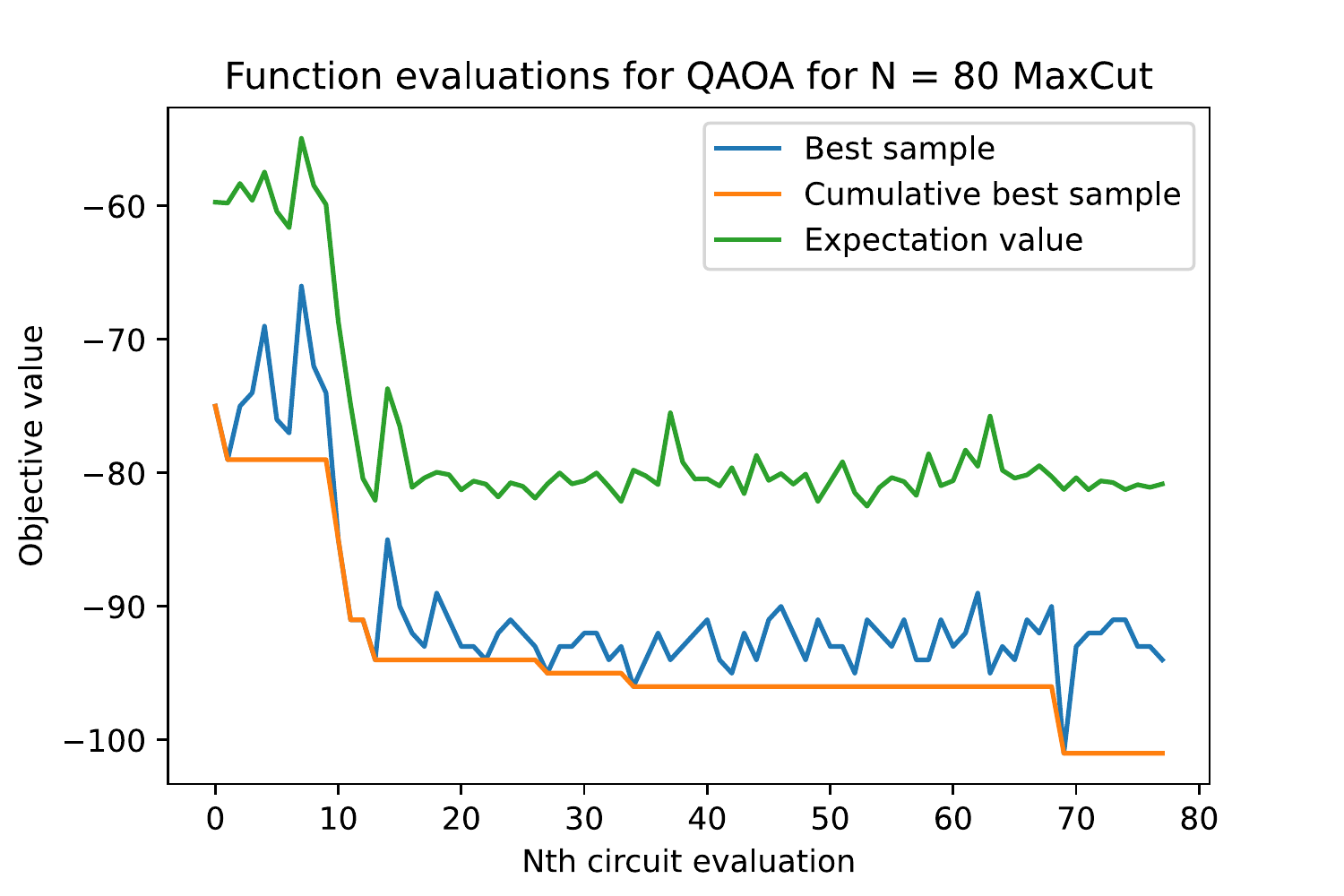}
    \caption{Full experimental data for all circuits submitted to the H1-1 machine. This data differs from Fig.~\ref{fig:exp_results} in that data for all circuits are displayed, not merely those that improve the objective.}
    \label{fig:second_exp_results}
\end{figure}

\begin{figure}[!t]
    \centering
    \includegraphics[width=0.5\textwidth]{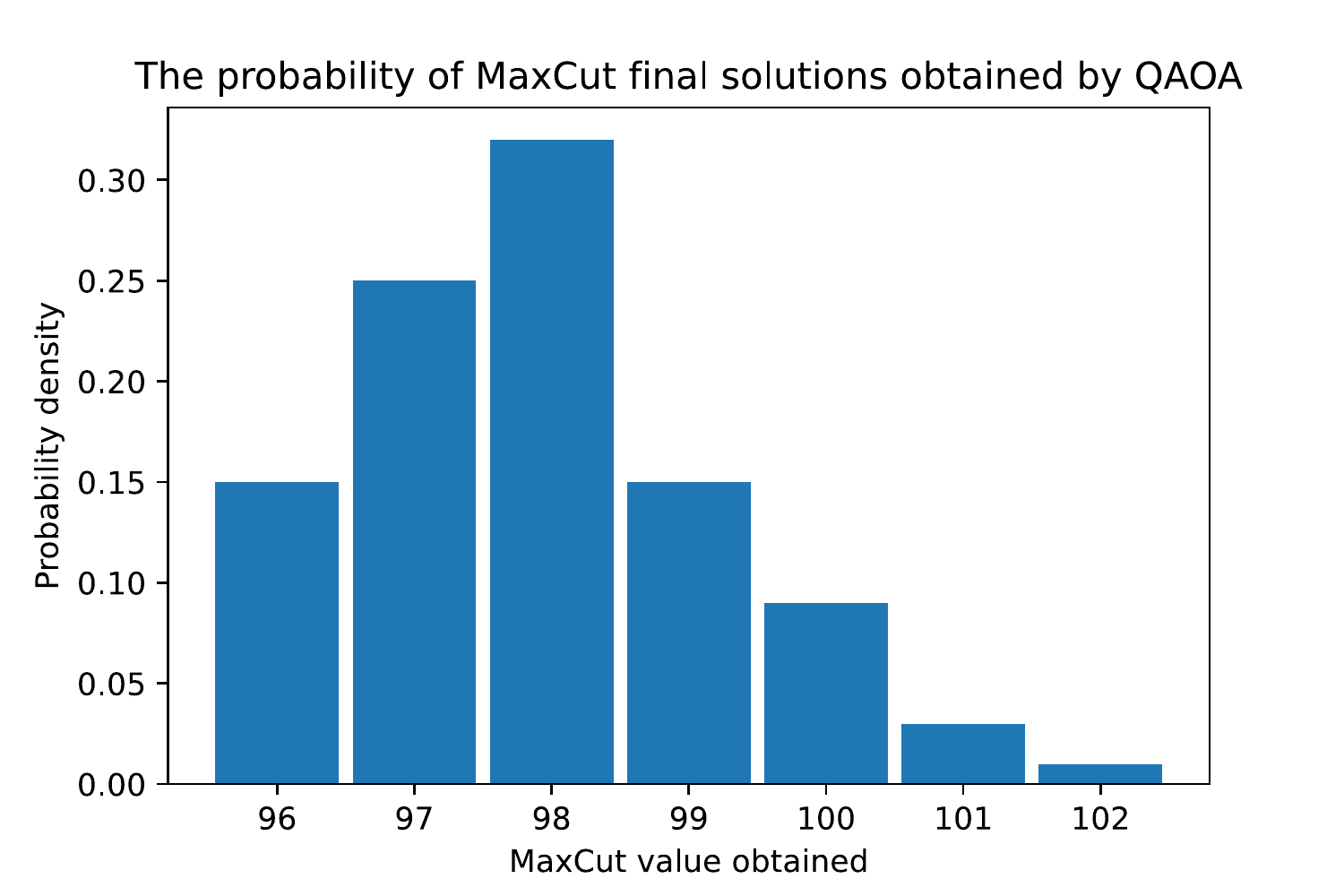}
    \caption{The probability of obtaining a particular cut value as the final QAOA solution, extracted from 100 circuits with 5000 shots each.}
    \label{fig:probability_cutval}
\end{figure}

\begin{figure*}[!t]
    \centering
    \includegraphics[width=\textwidth]{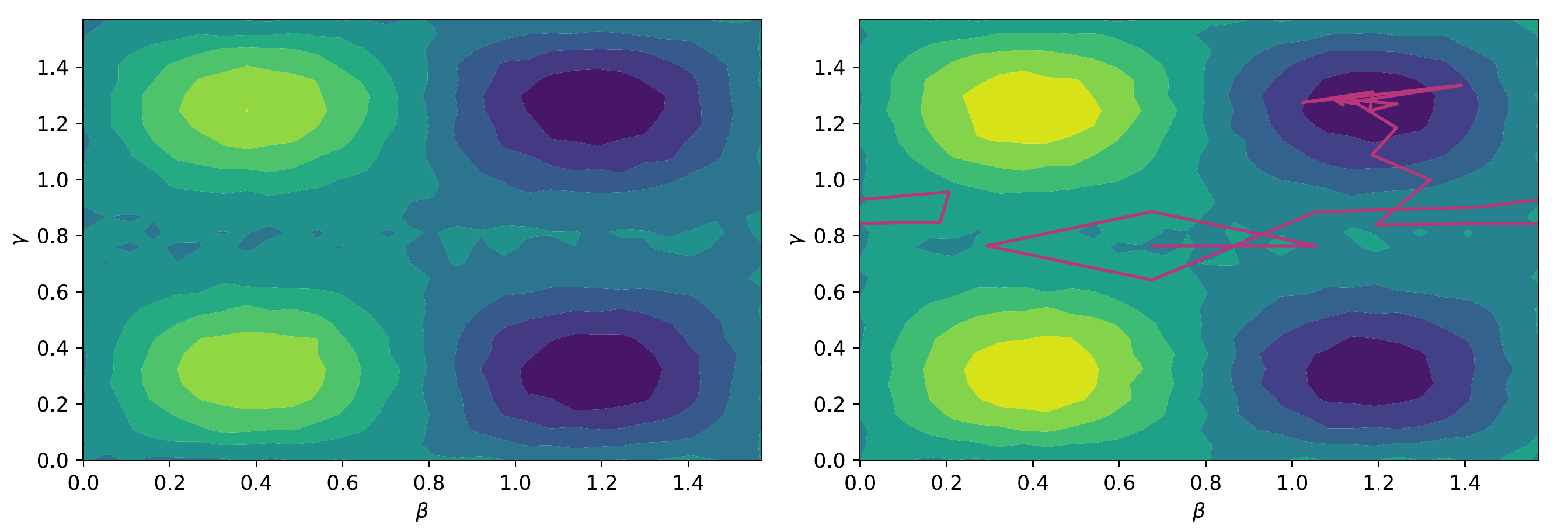}
    \caption{The energy landscape for the $N = 80$, $p=1$ MaxCut instance discussed in the main text as a function of $(\beta, \gamma)$, where the optimum corresponds to minimum energy (indigo). (Left) the noiseless landscape, generated using $30 \times 30 = 900$ ideal simulations of the QAOA circuit. (Right) the noisy landscape, generated using $30 \times 30 = 900$ simulations of the QAOA circuit with an appropriate error model for the Quantinuum H1-1 device. The actual optimization trajectory from the experimental data is overlaid in pink.}
    \label{fig:landscapes}
\end{figure*}

In Fig.~\ref{fig:second_exp_results} we display the full data taken from all circuits evaluated on the machine. After a short period of exploration in the parameter space, the optimizer converges rapidly around the tenth circuit evaluation. The subsequent circuits, during which the size of the trust region was decreased until meeting the convergence criterion of the optimizer, essentially end up providing extra samples at near-optimal parameters, increasing the probability of a high-quality solution (Fig.~\ref{fig:probability_cutval}). In Fig.~\ref{fig:landscapes} we confirm this interpretation with the full optimization trace overlaid on the energy landscape for this QAOA problem, generated by both noiseless and noisy simulation of the circuits over a lattice of parameter values.

The best cut value measured in any shot throughout the optimization procedure was 101. For comparison, the exact MaxCut value was computed to be 110 using the Gurobi solver \cite{gurobi} built into the Qiskit Optimization package \cite{Qiskit}. This exact solver took only $0.12$ seconds to execute on a laptop, unsurprisingly since $N=80$ is well within the purview of exact classical solvers. Given this exact value, the approximation ratio achieved by the QAOA run was $91.8\%$, which required about eleven hours of wall-clock time to execute on H1-1. In Fig.~\ref{fig:probability_cutval} we study the probability of obtaining a cut value of $101$ to assess how this value compares to the typical cut value that could be expected from a repetition of the experiment. As mentioned, from Fig.~\ref{fig:exp_results}, Fig.~\ref{fig:second_exp_results} and Fig.~\ref{fig:landscapes} it is apparent that the last $50$ circuits essentially serve only to provide extra samples at nearly-optimal parameters, for a total of $5000$ additional shots. Therefore, we take $100$ circuits evaluated with $5000$ shots each as a proxy for the full optimization procedure, and in Fig.~\ref{fig:probability_cutval} display the cut value obtained for the best shot for each circuit. It is clear that, although $101$ is slightly better than the average expected result, the distribution of results is fairly narrow and indeed the most likely cut value, $98$, is only marginally worse.

We also implemented the best-known classical approximation algorithm for MaxCut, the poly-time semidefinite-programming based Goemans-Williamson (GW) algorithm \cite{10.1145/227683.227684}, using the cvxopt solver \cite{cvxopt} in the PICOS Python API \cite{PICOS}. This algorithm provides a more scalable quantum-classical comparison, since at sufficiently large $N$ exact classical methods time out and one must rely on approximate algorithms to obtain the best possible solution in a reasonable time frame. The GW algorithm achieved a cut value of $104$, or an approximation ratio of $94.5\%$, and required $33$ seconds of computation time on a laptop. Note that when restricted to U3R graphs, the GW algorithm guarantees a lower bound for the approximation ratio of $93.26\%$ \cite{GW_approx_bound}.

\section{Discussion}

In this work we have presented a method (``qubit-reuse compilation") for executing quantum circuits using fewer qubits than naively required using mid-circuit measurements and resets. We detailed several algorithms that attempt to minimize the number of qubits required in the new circuit, and benchmarked their performance both numerically on MaxCut QAOA and analytically for certain structured circuits. Finally, we demonstrated the practical application and automated software implementation of these techniques by compressing and experimentally running the full optimization procedure for an $N = 80$, $p = 1$ MaxCut QAOA instance on the Quantinuum H1-1 device using $20$ qubits.

Since qubit reuse preserves the total number of gates in the original circuit, compressing a circuit using qubit-reuse compilation generally increases circuit depth. Consequently, some qubits in the compressed circuit may experience more memory errors by the time they are measured, since they may be active for the duration of a larger number of gates. Therefore, it is most likely beneficial to only compress circuits down to the number of qubits required to execute the circuit on a desired target device. A simple way to accomplish this when iteratively constructing a compiled circuit from a given measurement order is to prioritize using new qubits over reusing measured qubits at the beginning of the procedure, until the desired minimum number of qubits has been used, and has already been implemented in software.

It is important to note the sense in which a compressed circuit produced by qubit-reuse compilation is equivalent to the original circuit. Experiments performed on the compressed circuit (in the absence of errors) will produce bitstring samples from the same probability distribution as the original circuit. This is true even though the compressed circuit may have mid-circuit measurements, while the original circuit does not, and that the compressed circuit does not implement the same overall operator, and in fact acts on a Hilbert space of different dimension. While in a particular experimental realization a mid-circuit measurement collapses the quantum state to a particular quantum trajectory dependent on the measurement outcome, upon averaging over such realizations the estimated reduced density matrices on the measured qubits will be equivalent to the reduced density matrices produced by the original circuits. This is because before any qubit is measured in the compressed circuit, all of the operations in that qubit's causal cone are executed and therefore all of the necessary information for producing that qubit's reduced density matrix has been prepared. All of the other gates in the circuit not in this causal cone do not affect this qubit's density matrix and so can be delayed into a later section of the compressed circuit after this qubit has been reset and re-used. Another perspective of this explanation comes from considering the tensor network formulation of the circuit. In this case, the qubit-reuse compiled circuit can be obtained from the original circuit simply from diagrammatic manipulations of the tensor network reinterpreting the locations of input and output qubits, so they must sample from the same distributions.

Throughout this work we have assumed that the reader is generally interested in sampling the full output of a given quantum circuit. However, in many circumstances this is not the case and one is only interested in calculating a few-body correlation function of the output qubits. In this case, a further reduction in the number of qubits required to execute the circuit in order to measure this restricted set of output qubits may be possible. One simply constructs the modified circuit consisting of only the gates belonging to the causal cones of the desired output qubits, and utilizes the same algorithms for qubit reuse on this restricted circuit.

In many cases of interest, the original circuit is not unique but can be expressed using a different ordering and possibly different set of gates. One of the simplest possible examples of this occurs in both QAOA and in Hamiltonian simulation of Ising models, where many commuting $ZZ(\theta)$ gates are applied in a non-unique order that determines the resulting causal structure of the circuit. For this commuting gates problem, it is always possible to implement the block of commuting gates in a number of layers that is at most one more than the maximal degree of the interaction graph, by mapping the problem to edge coloring and using Vizing's theorem \cite{vizing, misra_gries}. Preliminary study indicates that minimizing the original circuit depth in this way also improves the amount of compression obtainable by qubit reuse, but in general the effects of the original circuit depth and gate ordering on causal structure merit further consideration.

\section{Acknowledgments}

The authors would like to thank many people within the broader Quantinuum team for their contributions towards this work. We acknowledge useful conversations with David Hayes, Steve Ragole, Jason Dominy, Mattia Fiorentini, David Amaro, and Marcello Benedetti. The experimental data in this work was produced by the Quantinuum H1-1 trapped ion quantum computer, powered by Honeywell, and we thank Brian Neyenhuis and the Quantinuum H1-1 operations team for their support and assistance.

\bibliography{References}
\bibliographystyle{apsrev}

\clearpage
\onecolumngrid

\section{Supplemental Materials}
\subsection{Constraints in the CP-SAT Model}

The CP-SAT model and accompanying constraints that are solved by the exact algorithm for qubit reuse are as follows:
 \begin{align*}
     \min C
 \end{align*}
 subject to
 
 \begin{Calignat}{3}
  & \forall t \in T, \nonumber \\
 &\qquad\qquad C \geq \sum_q c_{qt} \label{eq:first_constraint} \\
  & \forall t \in T,\: \forall q \in Q,\: \forall j \in C_q, \nonumber\\
  &\qquad\qquad m_{qt} = 1  && \hspace{-2.5em} \implies  \sum_{i=1}^t c_{ji} \geq 1 \label{eq:second_constraint}\\
  & \forall t \in T,\: \forall q \in Q, \nonumber \\[.1cm]
 &\qquad\qquad c_{q,t-1} = 1 && \hspace{-2.5em} \implies m_{q,t-1} + c_{q,t} = 1 \\[.2cm]
 & \forall t \in T,\: \forall q \in Q, \nonumber \\[.1cm]
 & \qquad\qquad m_{qt} = 1   &&\hspace{-2.5em} \implies  c_{qt} = 1 \\[.2cm]
  & \forall t \in T,\: \forall q \in Q, \nonumber \\
  & \qquad \qquad m_{qt} = 1  && \hspace{-2.5em} \implies \ \sum_{i = t+1}^N c_{qi} = 0\\
 & \forall q \in Q, \nonumber \\
 &  \qquad\qquad  \sum_t m_{qt} = 1 \\
 & \forall t \in T,\nonumber \\
 & \qquad \qquad \sum_q m_{qt}  = 1 \label{eq:last_constraint} 
 \end{Calignat}
 
 The numbering of constraints matches to that of the verbal description of the constraints in the main text.

\end{document}